\documentclass[10pt,journal]{IEEEtran}
\ifCLASSOPTIONcompsoc
  \usepackage[nocompress]{cite}
  \else
  \usepackage{cite}
\fi

\usepackage{graphicx} 
\ifCLASSINFOpdf
 \DeclareGraphicsExtensions{.pdf,.jpeg,.png}
\else
\fi
\usepackage{amsmath}
\usepackage{dsfont}
\usepackage{amsfonts}       % blackboard math symbols
\usepackage{amssymb}
\usepackage{mathtools}
\usepackage{amsthm}

\usepackage{hyperref}       % hyperlinks
\usepackage{url}            % simple URL typesetting
\usepackage{booktabs}       % professional-quality tables
\usepackage{nicefrac}       % compact symbols for 1/2, etc.
\usepackage{xspace}
\usepackage{xcolor,colortbl}
\usepackage{hhline}
\usepackage{bbm}
\usepackage[final]{changes}
\usepackage{comment}
\usepackage{tabularx}
\usepackage{multirow, makecell}
\usepackage{color,soul}
\usepackage{algorithm}
\usepackage{algorithmicx}
\usepackage{algpseudocode}

\newcommand{\E}{\mathrm{E}}

\def\dd{\mbox{d}}

\def\f{\frac}

\def\ie{\textit{i.e.}}
\def\eg{\textit{e.g.}}

\begin{document}

%\title{Optimal control of epidemics across networks using limited
%  testing and vaccination resources}

\title{Controlling epidemics through optimal allocation of test kits
  and vaccine doses across networks}

%\title{Optimal allocation of test kits and vaccine doses across
%  networks for controlling epidemics}

\author{Mingtao Xia, Lucas B\"{o}ttcher, Tom Chou%
  \IEEEcompsocitemizethanks{\IEEEcompsocthanksitem Mingtao Xia is in
    the Dept. of Mathematics at UCLA.\protect\\
E-mail: xiamingtao97@g.ucla.edu
\IEEEcompsocthanksitem Lucas B\"{o}ttcher is in the Dept.~of Computational Medicine at UCLA and at the Frankfurt School of Finance and Management. \protect\\E-mail: l.boettcher@fs.de
\IEEEcompsocthanksitem Tom Chou is in the Depts. of 
Computational Medicine and Mathematics at UCLA \protect\\E-mail: tomchou@ucla.edu}% <-this % stops an unwanted space

\thanks{Manuscript received July 18, 2021; revised August 31, 2021.}}

%{Shell \MakeLowercase{\textit{et al.}}: Bare Demo of IEEEtran.cls for Computer Society Journals}

%
%\author{Mingtao Xia}
%\email{xiamingtao97@ucla.edu}
%\affiliation{Dept.~of Mathematics, UCLA, Los Angeles, CA 90095-1555}
%\author{Lucas B\"{o}ttcher}
%\email{l.boettcher@fs.de}
%\affiliation{Dept.~of Computational Medicine, UCLA, Los Angeles, CA 90095-1766}
%\affiliation{Computational Social Science, Frankfurt School of Finance \& Management, Frankfurt am Main, 60322, Germany}
%\author{Tom Chou}
%\email{tomchou@ucla.edu}
%\affiliation{Dept.~of Computational Medicine, UCLA, Los Angeles, CA 90095-1766}
%\affiliation{Dept.~of Mathematics, UCLA, Los Angeles, CA 90095-1555}
%\date{\today}
%

\vspace{1cm}
\IEEEtitleabstractindextext{
\begin{abstract}
\added{Efficient testing and vaccination protocols are critical
  aspects of epidemic management. To study the optimal allocation of
  limited testing and vaccination resources in a heterogeneous contact
  network of interacting susceptible, recovered, and infected
  individuals, we present a degree-based testing and vaccination model
  for which we use control-theoretic methods to derive optimal testing
  and vaccination policies.  Within our framework, we find that
  optimal intervention policies first target high-degree nodes before
  shifting to lower-degree nodes in a time-dependent manner. Using
  such optimal policies, it is possible to delay outbreaks and reduce
  incidence rates to a greater extent than uniform and
  reinforcement-learning-based interventions, particularly on certain
  scale-free networks.}
\end{abstract}
%

\begin{IEEEkeywords}
Disease networks, epidemics, testing, vaccination, optimal control, reinforcement learning
\end{IEEEkeywords}
}

\maketitle

\IEEEdisplaynontitleabstractindextext

% \fi
\IEEEpeerreviewmaketitle

\vspace{9mm}

\IEEEraisesectionheading{\section{Introduction}\label{sec:introduction}}

Limiting the spread of novel pathogens such as SARS-CoV-2 requires
efficient testing~\cite{abdalhamid2020assessment,yelin2020evaluation}
and quarantine strategies~\cite{quilty2021quarantine}, especially when
vaccines are not available or effective. Even if effective vaccines
become available at scale, their population-wide distribution is a
complex and time-consuming endeavor, \added{influenced by, for example,
population age-structure \cite{vaccination_uk_2021,
  vaccination_age_1998, vaccination_age_2020}, vaccine hesitancy
\cite{vaccination_ramirez_2021}, and different objectives
\cite{vaccination_liceaga_2021}.}
%
%due to the special medical needs
%of immunocompromised patients, the mobility constraints of certain
%demographic groups, and other factors.

Until a sufficient level of immunity within a population is reached,
distancing and quarantine policies can also be used to help slow the
spread and evolutionary dynamics~\cite{CDC_newvariant} of infectious
diseases.  Epidemic modeling and control-theoretic approaches are
useful for identifying both efficient testing and vaccination
policies. For an epidemic model of SARS-CoV-2 transmission,
Pontryagin's maximum principle (PMP) has been used to derive optimal
distancing and testing strategies that minimize the number of COVID-19
cases and intervention costs~\cite{choi2021optimal}. Optimal control
theory has also been applied to a multi-objective control problem that
uses isolation and vaccination to limit epidemic size and
duration~\cite{bolzoni2019optimal}. Both of these recent
investigations describe the underlying infectious disease dynamics
through compartmental models without underlying network structure,
meaning that all interactions among different individuals are assumed
to be homogeneous. For a structured susceptible-infected-recovered
(SIR) model, optimal vaccination strategies have been derived for a
rapidly spreading disease in a highly mobile urban population using
PMP~\cite{ogren2000optimal}. Complementing these control-theory-based
interventions, a recent work \cite{wang2020risk} developed methods
relying on reinforcement learning (RL) to identify infectious
high-degree nodes (``superspreaders'') in temporal networks and reduce
the overall infection rate with limited medical resources. The
application of optimal control methods and PMP to a heterogeneous
node-based susceptible-infected-recovered-susceptible (SIRS) model
with applications to rumor spreading was studied
in~\cite{liu2020optimal}.

The machine-learning-based interventions of \cite{wang2020risk} showed
that RL is able to outperform intervention policies derived from
purely structural node characterizations that are, for instance, based
on centrality measures. However, the methods of \cite{wang2020risk}
were applied to rather small networks with a maximum number of nodes
of about 400. Here, we focus on a complementary approach by
formulating optimal control and RL-based target policies for a
degree-based epidemic model~\cite{newman2018networks} that is
constrained only by the maximum degree and not by the system size
(\ie, number of nodes). Early work by May and
Anderson~\cite{may1988transmission} employed such effective degree
models to study the population-level dynamics of human
immunodeficiency virus (HIV) infections. These degree-based models and
later
adaptations~\cite{barthelemy2004velocity,pastor2001epidemic,pastor2001epidemicPRL}
do not account for degree correlations. Effective degree models for
susceptible-infected-susceptible (SIS) dynamics with degree
correlations were derived in \cite{boguna2003absence} and applied to
SIR dynamics in \cite{barthelemy2005dynamical}. A further
generalization of these methods to model SIR dynamics with networked
and well-mixed transmission pathways was presented in
\cite{kiss2006effect}. For a detailed summary of degree-based epidemic
models, see~\cite{lindquist2011effective}.

In the next section, we propose and justify a degree-based epidemic,
testing, and quarantining model. An optimal control framework for this
model is presented in Sec.~\ref{sec:control_problem} and, given
limited testing resources, an optimal testing strategy is
calculated. We extend the same underlying disease model to include
vaccination in Sec.~\ref{sec:vaccination} and find optimal vaccination
strategies that minimize infection given a limited vaccination rate.
We summarize and discuss our results and how they depend on network
and dynamical features of the model in Sec.~\ref{sec:discussion}.  For
comparison, we also present in the Appendix a
reinforcement-learning-based algorithm that is able to approximate
optimal testing strategies for the model introduced in
Sec.~\ref{sec:model}.
\section{Degree-based epidemic and testing model}
\label{sec:model}
For the formulation of optimal testing policies that allocate testing
resources to different individuals in a contact network, we adopt an
effective degree model of SIR dynamics with testing in a static
network of $N$ nodes. Nodes represent individuals, and edges between
nodes represent corresponding contacts. Therefore, the degree of a
node represents the number of its contacts. If $K$ is the maximum
degree across all nodes, we can divide the population into $K$
distinct subpopulations, each of size $N_k$ ($k=1,2,\dots,K$) such
that all nodes in the $k^{\rm th}$ group have degree $k$. Therefore,
$N=\sum_{k=1}^{K}N_k$.

In our epidemic model, we distinguish between untested and tested
infected individuals.  Let $S_k(t)$, $I_k^{{\rm u}}(t)$,
$I_k^{{*}}(t)$, and $R_k(t)$ denote the numbers of susceptible,
untested infected, tested infected, and recovered nodes with degree
$k$ at time $t$, respectively. Since these subpopulations together
represent the entire population (the total number of nodes $N$), both
$N$ and $N_{k}$ are constants in our model. Their values satisfy the
normalization condition $S_k+I_k^{\rm u}+I_k^{*}+R_k=N_k$. The
corresponding fractions are
\begin{equation}
\begin{aligned}
s_k(t)=S_k(t)/N, & \quad i_k^{\rm u}(t)=I_k^{\rm u}(t)/N, \\
i_k^{*}(t)=I_k^{*}(t)/N, &  \quad r_k(t)=R_k(t)/N,
\end{aligned}
\end{equation}
such that $\sum_k (s_k+i_k^{\rm u}+i_k^{*}+r_k)=1$. Using an
effective-degree approach~\cite{may1988transmission,kiss2006effect},
we describe the evolution of the above subpopulations by
\begin{align}
\f{\dd s_k(t)}{\dd t} = & -k s_k(t)\sum_{\ell=1}^{K}{P(\ell\vert k) \over P(\ell)}
\big(\beta^{\rm u} i_{\ell}^{\rm u}(t) 
+ \beta^* i_{\ell}^{*}(t)\big),\label{eq:s_k}\\
\f{\dd i_k^{\rm u}(t)}{\dd t} = & k s_k(t)\sum_{\ell=1}^{K}
{P(\ell\vert k) \over P(\ell)}
 \big(\beta^{\rm u} i_{\ell}^{\rm u}(t)+
  \beta^* i_{\ell}^{*}(t)\big)\label{eq:i_k}\\[-4pt]
\: & \hspace{2.8cm}  - \gamma^u i_k^{\rm u}(t) -
\frac{f_k(t)}{N_k}i_k^{\rm u}(t),\nonumber\\
\f{\dd i_k^*(t)}{\dd t} = & - \gamma^* i_k^*(t) +
\frac{f_k(t)}{N_k} i_k^{\rm u}(t),\label{eq:i_k_ast}\\ 
\f{\dd r_k(t)}{\dd t} = &  \gamma^{\rm u} i_k^{\rm u}(t)+\gamma^*
i_k^*(t),\label{eq:r_k}
\end{align}
\added{where $P(\ell)=N_{\ell}/N$ is the degree distribution and
  $P(\ell\vert k)$ is the probability that a chosen node with degree
  $k$ is connected to a node with degree $\ell$. Our degree-based
  formulation of SIR dynamics with testing,
  Eqs.~\eqref{eq:s_k}--\eqref{eq:r_k}, is an approximation of the full
  node-based dynamics assuming that nodes of the same degree are
  equally likely to be infected at any given
  time~\cite{newman2018networks}.}

 \added{Susceptible individuals become infected through contact with
   untested and tested infected individuals at rates $\beta^{\rm u}$
   and $\beta^*$, respectively.  Untested and tested infected
   individuals recover at rates $\gamma^{\rm u}$ and $\gamma^*$,
   respectively. Differences in the recovery rates $\gamma^{\rm u}$
   and $\gamma^*$ reflect differences in disease severity of and
   treatment options for untested and tested infected
   individuals. Once recovered, individuals develop long-lasting
   immunity that protects them from reinfection. Temporary immunity
   can be easily modeled by using as SIS type model with or without
   delays.} A reduced transmissibility of tested infected (and
 potentially quarantined) individuals corresponds to setting $\beta^*
 \ll \beta^{\rm u}$.

\added{The total testing rate of nodes with degree $k$ is defined as
  $f_{k}(t)$, so that $f_{k}(t)\Delta t$ is the total number of tests
  given to all nodes with degree $k$ in time window $\Delta t$. Tests
  given to recovereds, susceptibles, and already-tested infecteds do
  not lead to quarantining and do not affect the disease
  dynamics. However a fraction $I_{k}^{\rm u}/(S_{k}+I_{k}^{\rm
    u}+I_{k}^{*}+R_{k}) \equiv I_{k}^{\rm u}/N_{k}$ of these
  $f_{k}(t)\Delta t$ tests will be administered to untested infecteds.
  Once infected nodes have been identified by testing, they can be
  quarantined and removed from the disease transmission dynamics.  If
  infected individuals who already have been tested strictly avoid
  future testing, more tests will be available for the other
  subpopulations, increasing the rate at which the remaining untested
  infecteds will be tested. In this case, the fraction of tests
  administered to untested infecteds is modified: $I_{k}^{\rm
    u}/(S_{k}+I_{k}^{\rm u}+R_{k}) \equiv I_{k}^{\rm
    u}/(N_{k}-I_{k}^{*})$. After normalizing by the total population
  $N$, we arrive at the testing terms $-f_{k}(t)i_{k}^{\rm u}/N_{k}$
  (Eqs.~\eqref{eq:s_k} and \eqref{eq:i_k}) or $-f_{k}(t)i_{k}^{\rm
    u}/[N_{k}(1-I_{k}^{\rm u}/N_{k})]$, respectively.}

\added{Biased testing can also be represented by using a testing
  fraction of the form $I_{k}^{\rm u}e^{b}/(I_{k}^{\rm
    u}e^{b}+S_{k}+I_{k}^{*}+R_{k})$, where $b>0$ increases the
  fraction of tests given to infecteds.  To correct for false positive
  tests, Eqs.~\eqref{eq:s_k}--\eqref{eq:r_k} can be modified by
  including an additional term that transfers the $I^{*}_k$ population
  back to $S_k$. False negatives can be accounted for by a reduction
  in $f_k(t)/N_k$.  For a detailed overview of statistical models that
  account for testing errors and bias, see
  \cite{bottcher2021using,testing_statistics21}.}

\begin{figure}
    \includegraphics[width=0.5\textwidth]{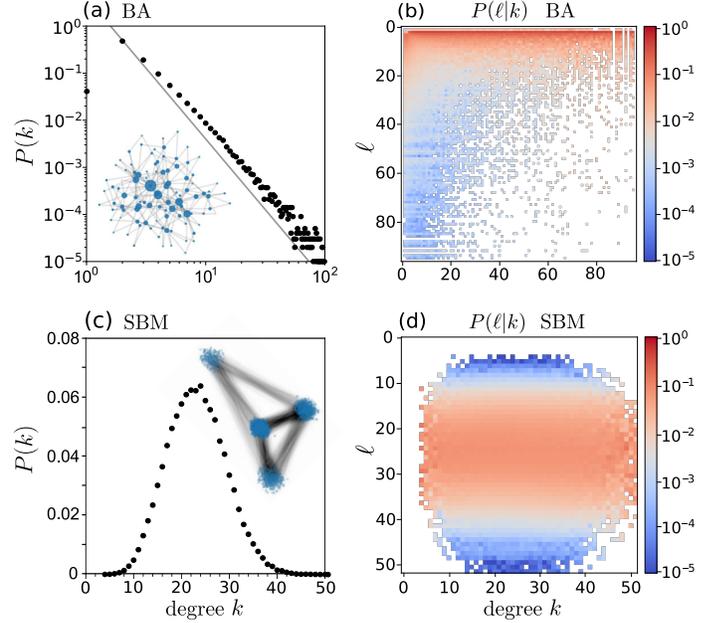}
\vspace{-5mm} 
\caption{Degree distribution of a Barab\'{a}si--Albert network and a
  stochastic block model. (a) The degree distribution of a
  Barab\'{a}si--Albert network with 99,939 nodes. Each new node is
  connected to $m=2$ existing nodes (\ie, the degree of each node is
  at least 2) using preferential attachment. Then nodes with degrees
  larger than 100~\cite{brown2013place} are removed from the network.
  The grey solid line is a guide-to-the-eye with slope
  -3~\cite{albert2002statistical}. The inset shows a realization of a
  Barab\'{a}si--Albert network with 100 nodes. Node size scales with
  their betweenness centrality. (b) The conditional probability
  $P(\ell\vert k)$ associated with the Barab\'{a}si--Albert network
  generated in (a).  (c) The degree distribution of a stochastic block
  model with four blocks and 100,000 nodes. The inset shows a
  realization of a stochastic block model with 800 nodes, but using
  the same block probability matrix. (d) The conditional probability
  $P(\ell\vert k)$ associated with the SBM. In both (b) and (d), all
  elements that are strictly zero are uncolored.}
    \label{fig:networks}
\end{figure}

\added{What remains is to assign network structures, extract
  $P(\ell\vert k)$ from them, and determine reasonable parameter
  values before calculating the optimal testing protocol $f_{k}(t)$.}
We apply our disease-control framework to (i) a Barab\'asi-Albert (BA)
network~\cite{barabasi1999emergence, albert2002statistical} and (ii) a
stochastic block model (SBM)~\cite{holland1983stochastic} with four
communities and probability matrix
\begin{equation}
P=10^{-4}\left(
\begin{array}{llll}
1&2&2&2\\
2&4&2&2\\
2&2&5&2\\
2&2&2&3
\end{array}
\right).
\end{equation}
\added{These two network types exhibit properties, such as hub nodes
  with high degrees and community structure, that are observable in
  real-world contact
  networks~\cite{brown2013place,zhao2012consistency}.  In the
  construction of the BA network, each new node is connected to 2
  existing nodes. Figure~\ref{fig:networks}(a) shows the degree
  distribution of a 99,939-node BA network that we use in this study.
  The conditional degree distribution $P(\ell|k)$ for a specific
  network can be directly evaluated as $E_{\ell, k}/(kN_{k})$ where
  $E_{\ell, k}$ is the number of edges connecting a node with degree
  $k$ with another node with degree $\ell$.  A heatmap of the
  conditional degree distribution matrix of the BA network with the
  degree distribution $P(k)$ shown in (a) is given in
  Fig.~\ref{fig:networks}(b).  The degree distribution and the
  conditional degree distribution matrix of the 100,000-node SBM
  network are shown in Figs.~\ref{fig:networks}(c) and (d),
  respectively. Taking into account empirical findings on the degree
  distributions in real-world contact networks~\cite{brown2013place},
  we use a degree cutoff of $k\leq K=100$.}

\added{Next, to constrain the parameter values, we first invoke
  estimates of the basic reproduction number (\ie, the average number
  of secondary cases that results from one case in a completely
  susceptible population), which for a network model is defined as
  \cite{van2002reproduction}}

%\cite{lindquist2011effective,lloyd2007network}}
%
\begin{equation}
    \mathcal{R}_0 = \rho(JV^{-1})
%
%\frac{\beta^{\text{u}}}{\beta^{\text{u}}
%+\gamma^{\text{u}}}\left[\langle k\rangle - 1 
%+ \frac{\text{Var}[k]}{\langle k \rangle}\right],
    \label{R0equation}
\end{equation}
\added{in which $\rho(\cdot)$ is largest eigenvalue (spectral radius),
  $V \equiv {\rm diag}(1/\gamma^{\rm u})\in\mathbb{R}^{K\times K}$ and
  $J\in\mathbb{R}^{K\times K}$ is the Jacobian of the linearized
  dynamical system (Eqs.~\eqref{eq:s_k} and \eqref{eq:i_k}) about the
  disease-free state with $f_{k} = 0$ corresponding to the initial,
  untested, and uncontrolled spread of the infection:}

\begin{equation}
J_{ij}= i P(j\vert i)\frac{N_i}{N_j}\beta^{\text{u}},\quad  i,j\leq K.
\label{F_next_gen_matrix}
\end{equation}
%
%
%$\langle k \rangle$ and $\text{Var}[k]$ denote the
  %mean degree and degree variance of the underlying network,
  %respectively.  
%
\added{This ``next generation'' method associates $\mathcal{R}_0$ with
  the largest eigenvalue inherent to the dynamical system. Additional
  expressions for $\mathcal{R}_0$ for an uncorrelated degree network
  are given in Appendix \ref{R0}.}

%For any general initial
%  perturbation (the degree of the initial infected node) that overlaps
%  with the eigenvector associated with the largest eigenvalue, the
%  $\mathcal{R}_0$ calculated \ref{R0equation} applies.}

\added{Empirically, the basic reproduction number for COVID-19 varies
  across different regions. For the early outbreak in Wuhan
  \cite{wang2020estimating}, $\mathcal{R}_0$ was estimated to be
  $3.49$, while for the early outbreak in Italy $\mathcal{R}_0\sim
  2.43 - 3.10$ \cite{d2020assessment}.  Here we set
  $\mathcal{R}_{0}=4.5$ which is suggested in \cite{katul2020global}
  as the reproduction number of the COVID-19 in early spreading
  without intervention measures. To find the proper value of
  transmissibility, we adjusted $\beta^{\rm u}$ until
  Eq.~\eqref{R0equation} yields $\mathcal{R}_{0}(\beta^{\rm
    u})=4.5$. Our source codes are publicly available at
  \url{https://gitlab.com/ComputationalScience/epidemic-control}.}
%
%Here we set $\mathcal{R}_0=2.91$ as reported
  %in \cite{lai2020severe}.  Using values of $\langle k \rangle$ and
 %$\text{Var}[k]$ from our generated networks in
 %Eq.~\eqref{R0equation}, we can estimate the prefactor
 %$\beta^{\text{u}}/(\beta^{\text{u}} +\gamma^{\text{u}})$.  The mean
 %recovery time since infection was found to be around 14 days for
 %different groups of people \cite{barman2020covid}, Therefore, we set
 %$\gamma^{\text{u}}$ to be $1/14(\text{day})^{-1}$ in our model.}% in
 %COVID-19 \cite{bottcher2020why}.}

%
\section{Allocating limited testing resources}
\label{sec:control_problem}
Without any testing constraints, it would be most effective for
disease control to use a testing budget $f_k(t)$ sufficiently large to
keep the fraction of untested individuals, $i_k^{\rm u}(t)$, close to
zero. In general, the testing budgets are constrained by

%If the testing resources are limited, the testing budgets
%$f_k(t)$ satisfy two constraints. First, the testing resources given
%to subpopulation with degree degree $k$ are at least $f_k^{\min}N_k$
%and do not exceed $f_k^{\max}N_k$
%
\begin{equation}
    f_k^{\min}\leq \frac{f_k(t)}{N_k}\leq f_k^{\max}, 
    \label{testlimit}
\end{equation} 
and the total testing rate is also bounded by availability and
logistics of testing $\sum_{k=1}^{K} f_k(t) = F(t)$.  The goal is to
determine, under these constraints, the function $f_k(t)$ or
$f_k(t)/N_k$ that most effectively reduces the total number of
infections. In practice, high-degree nodes (highly social individuals)
might be subject to more testing (and quarantining if positive) than
low-degree nodes because of their higher expected rate of infecting
others. This rationale would be translated as $f_k(t)/N_k >
f_{k'}(t)/N_{k'}$ if $k>k'$. In our numerical experiments, we use
sufficiently broad bounds of $f_k(t)$ and set $f_k^{\min}=f_{\min}$
and $f_k^{\max}=f_{\max}$.

To minimize the number of total infections over time, 
we define a loss function as
\begin{align}
L(T)=\int_0^{T}\!\!\mathrm{d}t\, \delta^{t}\!\sum_{k=1}^{K}ks_{k}(t)\!
\sum_{\ell=1}^{K}\!{P(\ell\vert k)\over P(\ell)}
\big(\beta^{\text{u}} i_{\ell}^{\text{u}}(t)
+ \beta^* i_{\ell}^*(t)\big),
\label{eq:loss}
\end{align}
where $\delta \in(0, 1]$ denotes a discount factor, which describes how
  we balance between minimizing current infections and future
  infections. For example, medical resources can better handle
  patients and new treatments can be given time to develop if the
  number of infections are spread over longer time periods. These
  effects can be effectively incorporated in the loss function by
  using $\delta < 1$. Minimizing the loss Eq.~\eqref{eq:loss} is
  equivalent to minimizing the number of infections, weighted by the
  discount factor $\delta^t$, in the time horizon $[0,T]$.

The associated Hamiltonian is
\begin{equation}
\begin{aligned}
H = & \delta^t\sum_{k=1}^{K}ks_k(t)\sum_{\ell=1}^{K} \! {P(\ell\vert k)\over P(\ell)}
\big(\beta^{\text{u}} i_{\ell}^{\text{u}}(t) + 
\beta^* i_{\ell}^*(t) \big) \\[-2pt]
\: & \hspace{6mm} + \sum_{k=1}^{K}\left(\lambda^s_k \f{\dd s_k(t)}{\dd t} 
+\lambda_k^{\text{u}}\f{\dd i^{\rm u}_k(t)}{\dd t} + 
\lambda^*_k \f{\dd i^*_k(t)}{\dd t}\right),
\end{aligned}
\end{equation}
where $\lambda^{s}_k$, $\lambda^{\rm u}_k$, and $\lambda^{*}_k$ are
adjoint variables associated with $s_k$, $i^{\rm u}_k$, and $i^*_k$,
respectively. The dynamics for $(\lambda_k^{s},\lambda_k^{\rm
  u},\lambda_k^*)$ obey
%
%\begin{equation}
\begin{align}
\f{\dd \lambda_k^s}{\dd t} = & \delta^t k\sum_{\ell=1}^{K} \! 
{P(\ell\vert k) \over P(\ell)}
\big(\beta^{\text{u}} i_{\ell}^{\text{u}}(t) 
+ \beta^* i_{\ell}^*(t)\big) \nonumber
 \\[-3pt]
\: & \quad  - \lambda_k^s k\sum_{\ell=1}^{K} \!{P(\ell\vert k) \over P(\ell)}
\big(\beta^{\text{u}} i_{\ell}^{\text{u}}(t)
+ \beta^* i_{\ell}^*(t)\big) \label{lambdas} \\[-2pt]
\: & \quad  + \lambda_{k}^{\rm u} k\sum_{\ell=1}^{K} \!{P(\ell\vert k) \over P(\ell)}
\big(\beta^{\text{u}} i_{\ell}^{\text{u}}(t) 
+ \beta^* i_{\ell}^*(t)\big), \nonumber \\
%
%\end{aligned}
%\end{equation}
%
%\begin{equation}
%\begin{aligned}
\f{\dd \lambda_k^{\text{u}}}{\dd t} = & \frac{\beta^{\text{u}}}{P(k)}
\sum_{j=1}^{K} P(k\vert j) s_{j}(t) (\delta^t-\lambda_{j}^{s} 
+ \lambda_{j}^{\text{u}}) \label{lambdaiu} \\[-5pt]
\: & \qquad\qquad\qquad - \gamma^{\rm u}\lambda_k^{\rm u}
- \frac{f_k(t)}{N_k}(\lambda_k^{\rm u} - \lambda_k^*),\nonumber \\
%
%\end{aligned}
%\end{equation}
%
%\begin{equation}
%\begin{aligned}
\f{\dd \lambda_k^*}{\dd t} = & \frac{\beta^{*}}{P(k)}
\sum_{j=1}^{K} \! P(k\vert j) s_{j}(t)
(\delta^t-\lambda_{j}^s + \lambda_{j}^{\text{u}})
- \gamma^*\lambda^*_k.
\label{lambdais}
\end{align}
%\end{equation}
%
\begin{figure*}[htb!]
\begin{center}
      \includegraphics[width=\textwidth]{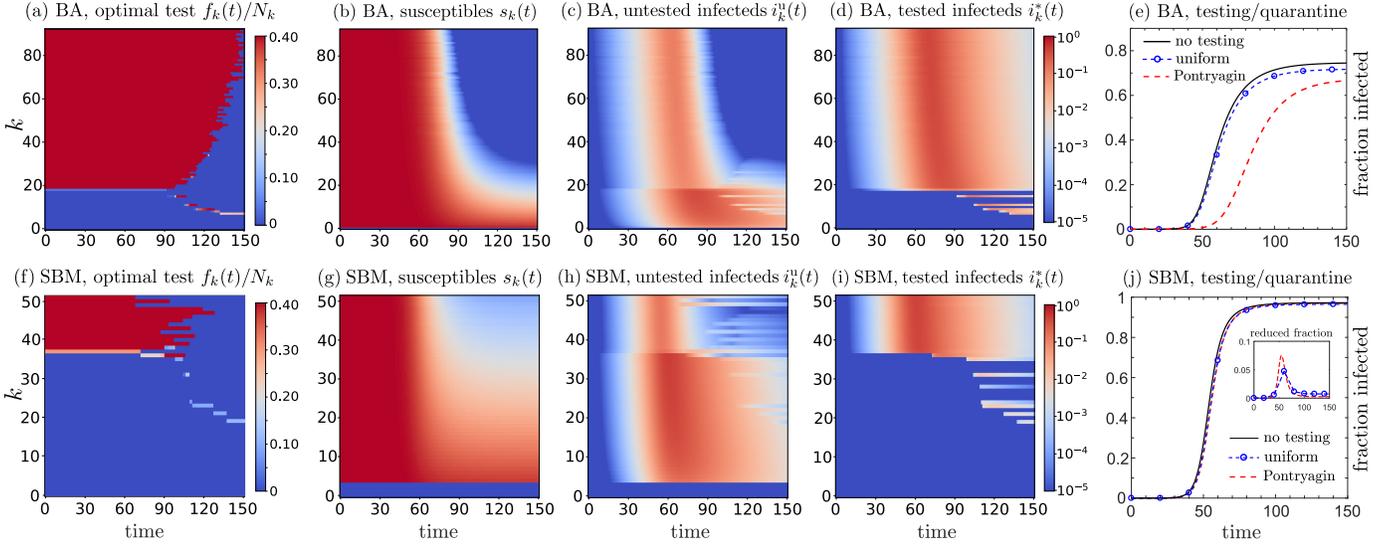}
	\end{center}
\vspace{-4mm}
\caption{\small \added{ Optimal testing and quarantining strategy for
    the BA network using $T=200$ and discount factor $\delta =0.95$.
    (a) A heatmap of the PMP-optimal testing strategy (see
    Alg.~\ref{algtesting}) for the BA network. The corresponding
    populations of degree-$k$ susceptibles, untested infecteds, and
    tested infecteds are plotted in (b-d), respectively. (e)
    Time-evolution of the total fraction infected
    $1-\sum_{k=1}^{K}s_{k}(t)$ under the PMP-optimal testing strategy
    (dashed red). The fractions infected under hypothetical uniform
    testing (dashed blue/circle) and no testing (black) scenarios are
    shown for comparison. For the BA network, optimal testing both
    delays and suppresses epidemic spreading more effectively than
    uniform testing. The bottom row (f-j) shows analogous results for
    the SBM network. (f-i) shows the corresponding optimal testing
    rates, susceptible, untested infected, and untested infected
    populations with degree $k$ as a function of time. (j) shows the
    fraction infected as a function of time. Although optimal testing
    and quarantining reduces the fraction infected relative to uniform
    or no testing, its effects are only modestly better. The effects
    of optimal testing strategies are greater in the BA network
    because its distribution of node degrees are more heterogeneous
    and testing and quarantining high-degree nodes can more
    effectively control disease spread. On the other hand, since the
    node degree distribution in the SBM network is sharply peaked, an
    optimal testing strategy is less effective overall.}}
     \label{fig70}
\end{figure*}
From the specific form 
\begin{equation}
\begin{aligned}
    H =  & \sum_{k=1}^{K}(\delta^t - \lambda_k^s+\lambda_k^{\text{u}})k\sum_{\ell=1}^{K}\!
{P(\ell\vert k) \over P(\ell)}
\big(\beta^{\text{u}} i_{\ell}^{\text{u}}(t)+ \beta^* i_{\ell}^*(t)\big)  \\[-3pt]
\: &\, \,   + \sum_{k=1}^{K} \Big[f_k(t)(\lambda_k^*-\lambda_k^{\text{u}})i_k^{\text{u}}(t) 
-\gamma^{\text{u}}i_k^{\text{u}}(t)\lambda_k^{\text{u}}
- \gamma^*i_k^*(t)\lambda_k^*\Big]
\end{aligned}
\end{equation}
and the constraint of the total budget $\sum_{k=1}^{K}f_k(t) = F(t)$, 
using Pontryagin's maximum principle, we calculate the optimal testing
rates according to $(f_k^*) = \mbox{argmin}_{f} H$.
To minimize $H$, we have to minimize the term
\begin{align}
    \sum_{k=1}^{K} \frac{f_k(t)}{N_k}(\lambda_k^*-\lambda_k^{\text{u}})i_k^{\text{u}}(t).
    \label{PMminitest}
\end{align}
Hence, we have to maximize those $f_k(t)$ with smallest coefficients
$(\lambda_k^*-\lambda_k^{\text{u}})i_k^{\text{u}}(t)/N_{k}$ and
minimize those $f_k(t)$ with largest coefficients, given the total
budget constraint. In other words, we should give testing resources to
those groups presumed to be at the highest risk, as quantified by the
quantity $(\lambda_k^*-\lambda_k^{\text{u}})i_k^{\text{u}}(t)/N_{k}$.

We use the PMP-based testing algorithm outlined in Appendix \ref{ALGO}
to iteratively calculate the optimal testing strategy and loss
function \eqref{eq:loss}.  Numerical experiments for a BA and an SBM
network were performed using a degree cutoff $K=100$ (see
Sec.~\ref{sec:model}). In accordance with empirical data on COVID-19
patients~\cite{barman2020covid,
  wilasang2020reduction,hyafil2020analysis}, we set
$\gamma=\gamma^{\text{u}}=\gamma^{*}=(14)^{-1}/\text{day}$ and
$\beta^{*}=\beta^{\text{u}}/10$. The transmissibility of untested
individuals, $\beta^{\text{u}}$, is calculated according to
Eq.~\eqref{R0equation} as $\beta^{\text{u}}=0.0417/\text{day}$ for the
BA network and $\beta^{\text{u}}=0.0130/\text{day}$ for the SBM
network.  We set the discount factor $\delta=0.95$ so that initial
infections contribute more to the loss function \eqref{eq:loss}.  The
total daily number of SARS-CoV-2 tests in the US after an initial
ramping-up phase in 2020 is about
0.6\%/\text{day}~\cite{testing_statistics21}. Hence, we set
\begin{equation}
    \sum_k f_k(t) = 0.006N,
    \label{testuniform}
\end{equation}
and $f_{\min} = 0, f_{\max}=0.4N_k$. As initial condition, we use
\begin{equation}
\begin{aligned}
~s_k(0) = P(k)-i_k^{\text{u}}(0), & \,\, ~i_k^*(0)=0, \\
i_k^{\text{u}}(0)=10^{-6}P(k),\quad  & \,\, ~r_k(0) = 0,
\end{aligned}
\label{ICs}
\end{equation}
\added{corresponding to about 0.1 of an infected individual uniformly
  distributed on $N \approx 10^{5}$ susceptible nodes. The optimal
  testing strategy is supposed to identify those nodes that are most
  likely to be infected and transmit the disease to others. Upon using
  $T=200, \Delta{t}=0.1$ and $\delta=0.95$, we find the optimal
  testing strategy $f_{k}(t)/N_{k}$ for our BA network and plot it in
  Fig.~\ref{fig70}(a). Here Eqs.~\eqref{eq:s_k}--\eqref{eq:r_k} and
  \eqref{lambdas}--\eqref{lambdais} are solved using an improved Euler
  method. For the BA network, the value of the loss function defined
  in Eq.~\eqref{eq:loss} is $L(T=200)=0.0114$ under the optimal
  testing strategy, while it is $L(T=200)=0.0330$ under uniform
  testing}
\begin{equation}
f_k=F_0\frac{N_k}{N}.
\label{uniftest}
\end{equation}  
\added{Figs.~\ref{fig70}(b-d) show the associated
  populations under optimal testing, while (e) shows the dynamics
  of the fraction of nodes infected, $1-\sum_{k=1}^{K}s_{k}(t)$, is
  significantly slowed relative to the no testing (black) and uniform
  testing  (dashed blue/circle) cases.  Fig.~\ref{fig70}(f)
  plots the optimal testing rate for the SBM network. (g-i) show the
  corresponding subpopulations, and (j) plots the fraction of nodes
  infected under PMP-optimal, uniform, and no-testing conditions. For
  the SBM network, $L(T=200)=0.0564$ under the optimal strategy and
  $L(T=200)=0.0571$ under the uniform testing strategy, suggesting
  that the PMP approach yields better solutions than uniform testing.
  However, the improvement is modest and the SBM network is rather
  insensitive to testing and quarantining. The slight improvement from
  testing is shown by the \textit{reduction} in the fraction infected
  relative to the no testing case (inset).}

%{\color{red}...and suppress initial infections for the SBM network
%  because later infections have a smaller weight $\delta^t$ in the loss
%  function Eq.~\eqref{eq:loss}???}

\added{In both networks, nodes with larger degrees are more likely to be
tested at the beginning of the outbreak [Figs.~\ref{fig70}(a,f)],
indicating that people with more contacts are more likely to infect
others or get infected and should be given priority to get
tested. Yet, in both networks, as time evolves, the optimal testing
strategy tends to shift focus from higher degree nodes to nodes with
smaller degrees because testing those nodes that were infected and
have already recovered is not meaningful in terms of disease
control.}

\added{Comparing Figs.~\ref{fig70}(e) and (j), we see that the
  differences between optimal and uniform testing are larger for the
  BA network compared to the SBM.  A possible explanation for this
  behavior is that in the BA network, the degree distribution $P(k)$
  decays algebraically. Therefore, as long as testing focuses
  primarily on high-degree nodes, the spreading of the disease can be
  controlled very effectively since the majority of nodes have low
  degree and are more unlikely to be infected. On the other hand, for
  our SBM network, the degrees of most nodes are close to each other
  and larger than 10, indicating that nodes with a small degree are
  more likely to be infected compared to the BA network. Even if we
  use the same uniform testing rates [see Eq.~\eqref{uniftest}] in
  both networks, the proportion of infections in the BA network is
  less than that in the SBM network. Nodes with small degrees in the SBM
  are more likely to be infected than those in the BA network because
  they are more connected to other nodes. 
  }
%

%%%%%%%%%%%%%%%%%%%%%%%%%%%%%%%%%%%%%%%%%%%%%%%%%%%%%%%%%%%%%%%%%
%%%%%%%%%%%%%%%%%%%%%%%%%%%%%%%%%%%%%%%%%%%%%%%%%%%%%%%%%%%%%%%%%
\section{Optimal vaccination policy}
\label{sec:vaccination}
%%%%%%%%%%%%%%%%%%%%%%%%%%%%%%%%%%%%%%%%%%%%%%%%%%%%%%%%%%%%%%%%%
%%%%%%%%%%%%%%%%%%%%%%%%%%%%%%%%%%%%%%%%%%%%%%%%%%%%%%%%%%%%%%%%%

\added{Optimal vaccination has also been studied within the classic
  SIR model \cite{ZAMAN_2017}. However, devising vaccination
  strategies based on social network structure may provide a more
  refined and efficient way of administering vaccines and
  extinguishing an epidemic. Our simple testing model presented in the
  previous section can be straightforwardly adapted to describe
  vaccination on a network. The goal is to determine the optimal
  allocation of vaccine doses to a population with heterogeneous
  contacts to minimize the impact of the infection across the entire
  population.}

For COVID-19, there are a variety of vaccines that require one or two
shots~\cite{peiffer2021covid}. In our simulations, we assume that the
administered vaccine provides full protection after one shot and that
a vaccinated individual will instantly leave the susceptible group and
enter the recovered group. This means that vaccinated individuals will
no longer be infectious and can be treated as ``recovered'' after
receiving one vaccination dose. Other mechanisms such as prime-boost
protocols and time delays between vaccination and onset of immune
response can also be accounted for in similar models as detailed in
\cite{bottcher2021decisive}.

We reformulate Eqs.~\eqref{eq:s_k}-\eqref{eq:r_k} to study optimal
vaccination protocols that are constrained by vaccine supplies in
a heterogeneous population. For simplicity, we do not take into
account the effect of testing and quarantining when devising optimal
vaccinating strategies, although testing and vaccination can be
performed concurrently. The resulting rate equations are
\begin{align}
\f{\dd s_k(t)}{\dd t} & = - \beta ks_k(t)\sum_{\ell=1}^{K}\!
\frac{P(\ell\vert k)}{P(\ell)} i_{\ell}(t) 
- \frac{v_k(t)}{N},\label{vaccines}\\
\f{\dd i_k(t)}{\dd t} & = \beta k s_k(t)\sum_{\ell=1}^{K}\!
\frac{P(\ell\vert k)}{P(\ell)}
i_{\ell}(t) - \gamma i_k(t),\label{vaccinei}\\
\f{\dd r_k(t)}{\dd t} & = \gamma i_k(t) +\frac{v_k(t)}{N},\label{vacciner}
\end{align}
where $v_k(t)$ is the \added{rate of vaccination of susceptibles with
  degree $k$ at time $t$. Once vaccinated, susceptibles become
  ``recovered'' because they are immunized and no longer susceptible
  to the infection. The total rate of administering vaccines at time
  $t$ is defined as}
\begin{align}
    \sum_{k=1}^{K}v_k(t) = V(t).
    \label{constraintvaccine}
\end{align}
\added{In other words, in time increment $\Delta t$ at time $t$, we
  can administer only $V(t)\Delta t$ doses. Eq.~\eqref{vaccines}
  assumes that vaccination is resource-limited and that the rate of
  protecting susceptibles is proportional only to the rate $v_{k}(t)$
  of administering vaccines.} In addition, we assume that the
vaccination rate for different subpopulations is confined to the
interval
\begin{equation}
    v_{\min}\leq \frac{v_k(t)}{Ns_k(t)} \leq v_{\max},
    \label{vaclimit}
\end{equation}
where $v_{\min}, v_{\max}\in[0, 1]/\textrm{day}$ are minimum and
maximum vaccination rates. \added{Note that vaccines are allocated
  only to susceptibles, while tests are typically given to individuals
  of all categories: susceptible, infected, and recovered, according
  to their relative proportions.}  To formulate the vaccine
distribution problem in a heterogeneous contact network, we use the
following loss function
\begin{align}
    L(T)=\int_0^{T}\!\mathrm{d}t\,  \delta^t\sum_{k=1}^{K}ks_k(t)
\sum_{\ell=1}^{K}\!{P(\ell\vert k)\over P(\ell)} 
\beta(t) i_{\ell}(t),
    \label{Lossvaccine}
\end{align}
with the aim of minimizing the total number of infections over time
(with a constant discount factor $\delta\in(0,1]$) by appropriately
  distributing vaccines to groups with different degree $k$ at
  different rates.
\begin{figure*}[htb!]
        \begin{center}
    \includegraphics[width=0.93\textwidth]{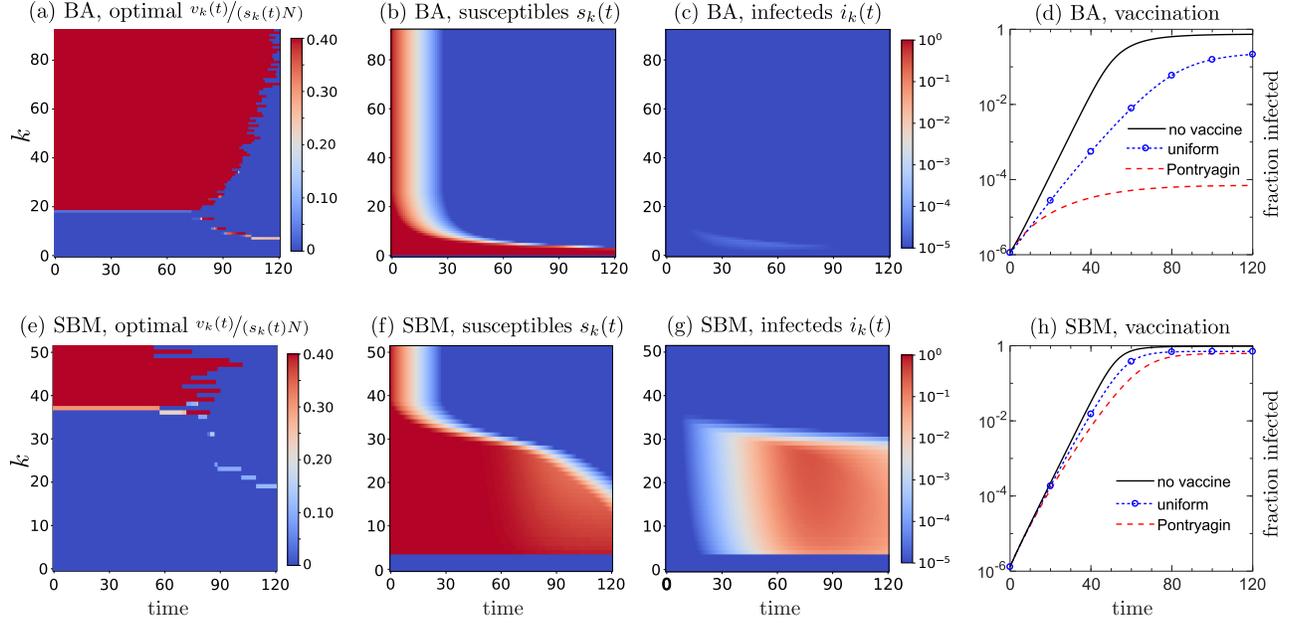}
	\end{center}
\vspace{-3mm}
\caption{\small \added{Optimized vaccination model.  (a) Heatmap of
    the optimal vaccination strategy $v_{k}(t)/(s_k(t)N_{k})$ for the
    BA network given by Alg.~\ref{algtesting}. (b,c) show the
    corresponding susceptible and infected subpopulations $s_{k}(t)$
    and $i_{k}(t)$, while (d) plots the fraction infected as a
    function of time, derived from solving
    Eqs.~\eqref{vaccines}--\eqref{vacciner} under optimal vaccination
    using a discount factor $\delta=0.95$.  The dashed red curve
    indicates the fraction infected under optimal vaccination.  For
    comparison, the infected population under no vaccination (solid
    black) and constant, uniform (dashed blue/circles) vaccination are
    also plotted and show how optimizing vaccination significantly
    suppresses infectivity. (e-h) show the corresponding quantities
    for the SBM network. Optimal vaccination is less effective at
    decreasing infection in the SBM network than in the BA network,
    again because of the SBM's peaked (more homogeneous) node degree
    distribution. Note from the logarithmic scale that vaccination is
    qualitatively more effective in reducing infections than testing
    and quarantining.}}
     \label{figvac}
\end{figure*}

To minimize the loss function \eqref{Lossvaccine}, we construct
the Hamiltonian
\begin{equation}
\begin{aligned}
H = & \beta\delta^t\sum_{k=1}^{K}ks_k(t)\sum_{\ell=1}^{K}
{P(\ell\vert k)\over P(\ell)}i_{\ell}(t) \\[-2pt]
\: & \qquad\qquad\qquad  + \sum_{k=1}^{K}\Big(\lambda^{s}_k \f{\dd s_k(t)}{\dd t} +\lambda_k^i
\f{\dd i_k(t)}{\dd t}\Big), \\
\: = & \beta \sum_{k=1}^{K}(\delta^t - \lambda_k^{s}+\lambda_k^{i})k s_k(t)
\sum_{\ell=1}^{K}\! {P(\ell\vert k)\over P(\ell)} i_{\ell}(t)  \\[-2pt]
\: & \qquad\qquad\qquad  + \sum_{k=1}^{K}\Big(\frac{v_k(t)}{N}\lambda_k^{s}(t) -\gamma i_k(t)\Big)
\label{Hamiltonvaccine}
\end{aligned}
\end{equation}
where $\lambda_k^{s}$ and $\lambda_k^i$ are the Lagrange multipliers
satisfying the differential equations
\begin{align}
\f{\dd \lambda_k^s}{\dd t} = & \beta k \sum_{\ell=1}^{K}{P(\ell\vert k)\over P(\ell)} 
i_{\ell}(t)
(\delta^{t} -\lambda_k^{s}+ \lambda_{k}^{i})  \label{adjointvacs}\\
\f{\dd \lambda_k^i}{\dd t} = & \frac{\beta}{P(k)}\sum_{j=1}^{K}P(k\vert j) 
 j  s_{\ell}(t) (\delta^t-\lambda_{j}^s 
+ \lambda_{j}^{i})  - \gamma\lambda_k^s.\label{adjointvac}
\end{align}

Therefore, minimizing the loss function \eqref{Lossvaccine} for the
given dynamics is equivalent to minimizing the Hamiltonian
\eqref{Hamiltonvaccine} using Pontryagin's maximum principle. From the
constraints \eqref{constraintvaccine} and \eqref{vaclimit}, minimizing
the Hamiltonian is achieved by giving vaccination $v_k$ to those
subpopulations with the smallest $\lambda_k^s$. We can still use
Alg.~\ref{algtesting} to solve the minimization problem
\eqref{Lossvaccine} numerically and obtain the optimal strategy.

In the US, about two million doses of SARS-CoV-2 vaccines were
delivered in May 2021~\cite{mathieu2021global}, most of which were
two-dose vaccines. Since approximately $0.3\%$ of the entire US
population is fully vaccinated daily, we set $V(t)=0.003N/\text{day}$,
$v_{\min}=0/\text{day}$, and $v_{\max}=0.4/\text{day}$ in the
constraint \eqref{vaclimit}. The infection rates $\beta$ are set to
be $0.0417/\text{day}$ for the BA network and $0.0130/\text{day}$ for
the SBM network, and the recovery rate $\gamma=(14)^{-1}/\text{day}$. For
comparison, we also simulate a vaccination strategy with a uniform
vaccination rate
\begin{equation}
    v_k(t) = \frac{s_k(t)V(t)}{\sum_{k=1}^{K} s_k(t)}.
    \label{vaccineunif}
\end{equation}
In all simulations, we use the following initial condition:
\begin{equation}
i_k(0)=10^{-6}P(k), ~r_k(0) = 0,s_k(0) = P(k)-i_k(0).
\label{ICv}
\end{equation}
\added{We plot the PMP-optimal vaccination strategy $v_{k}/N_{k}$ in
  Figure~\ref{figvac}(a) and the corresponding susceptible and
  infected $k$-degree subpopulations $s_{k}(t)$ and $i_{k}(t)$ in (b)
  and (c).  We set $T=150, \Delta{t}=0.1$ and we use an improved Euler
  method to numerically solve Eqs.~\eqref{vaccines}--\eqref{vacciner},
  \eqref{adjointvacs}--\eqref{adjointvac}. Alg.~\ref{algtesting} is
  applied (without the infected and tested compartment) to determine
  the optimal vaccination strategy by the PMP approach. For the BA
  network, $L(T=150)=1.241\times 10^{-5}$ under the PMP-optimal
  strategy and $L(T=150)=0.01990$ under a uniform vaccination
  rate. Figure~\ref{figvac}(d) shows that the optimal vaccination
  strategy on a BA network significantly reduces the fraction infected
  compared to the uniform vaccination strategy. (e-h) show the
  corresponding quantities for the SBM network for which
  $L(T=150)=0.0211$ under the optimal vaccination strategy and
  $L(T=150)=0.0360$ under a constant, uniform vaccination strategy. In
  both networks, the optimal vaccination strategies obtained via
  Alg.~\ref{algtesting} tend to prioritize those nodes with higher
  degrees first and eventually expand to those nodes with smaller
  degrees [see Figs.~\ref{figvac}(a) and (e)].  As with testing and
  quarantining, the reduction in the fraction infected by vaccination
  is greater in the BA network.  Since the BA network has a degree
  distribution with algebraic decay, the effect of the optimal
  vaccination strategy will be more pronounced than for the SBM, whose
  nodes have similar degrees.}
\section{Discussion and Conclusions}
\label{sec:discussion}
Effective testing and vaccination strategies are an essential part of
epidemic management. In this paper, we derived optimal testing and
vaccination policies by applying Pontryagin's maximum principle to a
degree-based epidemic model in a heterogeneous contact network. We
complemented our analytical results with reinforcement learning (RL)
approaches that identify effective policies. (see Appendix~\ref{sec:rl})
\begin{figure*}[htb!]
\begin{center}
 \includegraphics[width=0.94\textwidth]{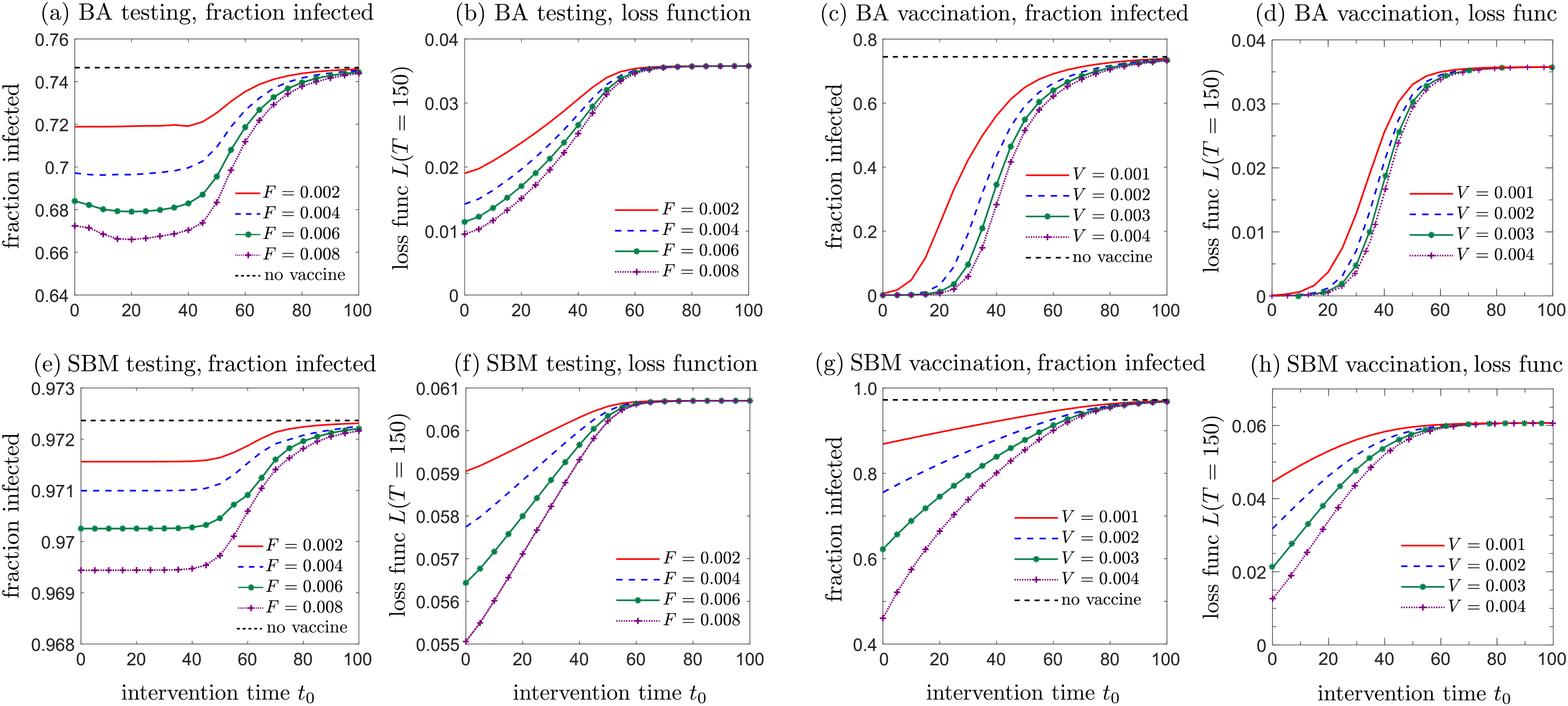}
\end{center}
\vspace{-3mm}
\caption{\small \added{Total fraction infected under testing or
    vaccination model as a function of different intervention starting
    times $t_{0}$. We minimize the corresponding loss function at
    $T=150$ and use $\delta=0.95$.  (a) The fraction infected in the
    BA network as a function of start times using different testing
    amplitudes $F$. At larger $F$, there is a \textit{decrease} in
    infected population for \textit{later} starts of testing at early
    times. This nonmonotonic response arises from $\delta < 1$ that
    weights early infections more strongly in the loss function, which
    remains monotonic in $t_{0}$ under optimization as shown in
    (b). The effect of delayed vaccination on the fraction infected is
    shown in (c), with the corresponding loss function shown in (d).
    For the SBM network, the fraction infected as a function of
    testing start time shown in (e) reflects the small effect of
    testing on the infected population. Both the fraction infected and
    loss function (f) are monotonic in starting time. The starting
    time dependence of the fraction infected on an optimally
    vaccinated SBM network in (g) shows a monotonic and smooth
    decrease in effectiveness as vaccination is delayed. In (h), the
    loss function for vaccination on the SBM network also
    monotonically increases with start time.}}
\label{fig:timing}
\end{figure*}

Our analytical results show that optimal testing and vaccination
policies under resource constraints initially tend to prioritize nodes
with higher degrees to control spread of the disease.  In situations
where the number of contacts of individuals is known or can be
estimated with reasonable precision, Algs.~\ref{algtesting} and
\ref{algqlearning} may be useful to identify effective epidemic
management strategies. \added{Using our control-theoretic approach, we
  also explored the relative effectiveness of testing and vaccination
  under different conditions.}

\subsection{\added{Effects of delayed intervention}}

\added{First, we consider the effectiveness of interventions as a
  function of the time between the first infection and the
  implementation of testing or vaccination. The initial conditions are
  set to be the same as Eqs.~\eqref{ICs} and \eqref{ICv}.
  Fig.~\ref{fig:timing} shows the total fraction infected and the loss
  functions at $T=150$, for both the BA and SBM networks, as a
  function of intervention starting time $t_{0}$. We set $F =
  F(t)\mathds{I}_{t>t_0}$ or $V = V(t)\mathds{I}_{t>t_0}$ and explore
  the effects of different constant levels of test kits or vaccine
  availability, $F(t)=0.002, 0.004, 0.006, 0.008 N/\text{day}$ and
  $V(t)=0.001, 0.002, 0.003, 0.004N/\text{day}$, respectively. The
  transmissibility rates $\beta^{\text{u}}, \beta^*, \beta$ and the
  recovery rates $\gamma^{\text{u}}, \gamma^*, \gamma$ are set to the
  same values as those used in Section \ref{sec:control_problem} for
  the testing model and those used in Section \ref{sec:vaccination}
  for the vaccination model.}

\added{At higher levels of $F$ and $V$, the high-$k$ nodes are
  addressed sooner and total infections can be reduced. For the
  vaccination model applied to both networks, an earlier intervention
  time will always lead to fewer infected nodes.  In the BA network,
  there exists an overall vaccination-rate-dependent starting time
  before which disease spread can be totally suppressed.  Overall, we
  found that earlier and stronger intervention measures lead to more
  effective control of the spreading of the disease and a smaller loss
  function defined in Eqs.~\eqref{eq:loss}, \eqref{Lossvaccine}.}

\added{However, for $\delta<1$ ($\delta = 0.95$ in this study), we found
  that the final infected proportions can actually \textit{decrease}
  with later testing starting times $t_{0}$ shortly after the initial
  infection, particularly with larger $F$. The testing loss function
  monotonically increases with $t_{0}$, a feature that is not
  preserved in the total final infection ratio. This qualitative
  difference can arise when $\delta < 1$ because the strategy of earlier
  testing tends to minimize initial infections in a way that reduces
  the loss function, even though the corresponding infection levels
  may be even larger than those associated with later starts in testing.}
\begin{figure*}[htb!]
\begin{center}
 \includegraphics[width=0.95\textwidth]{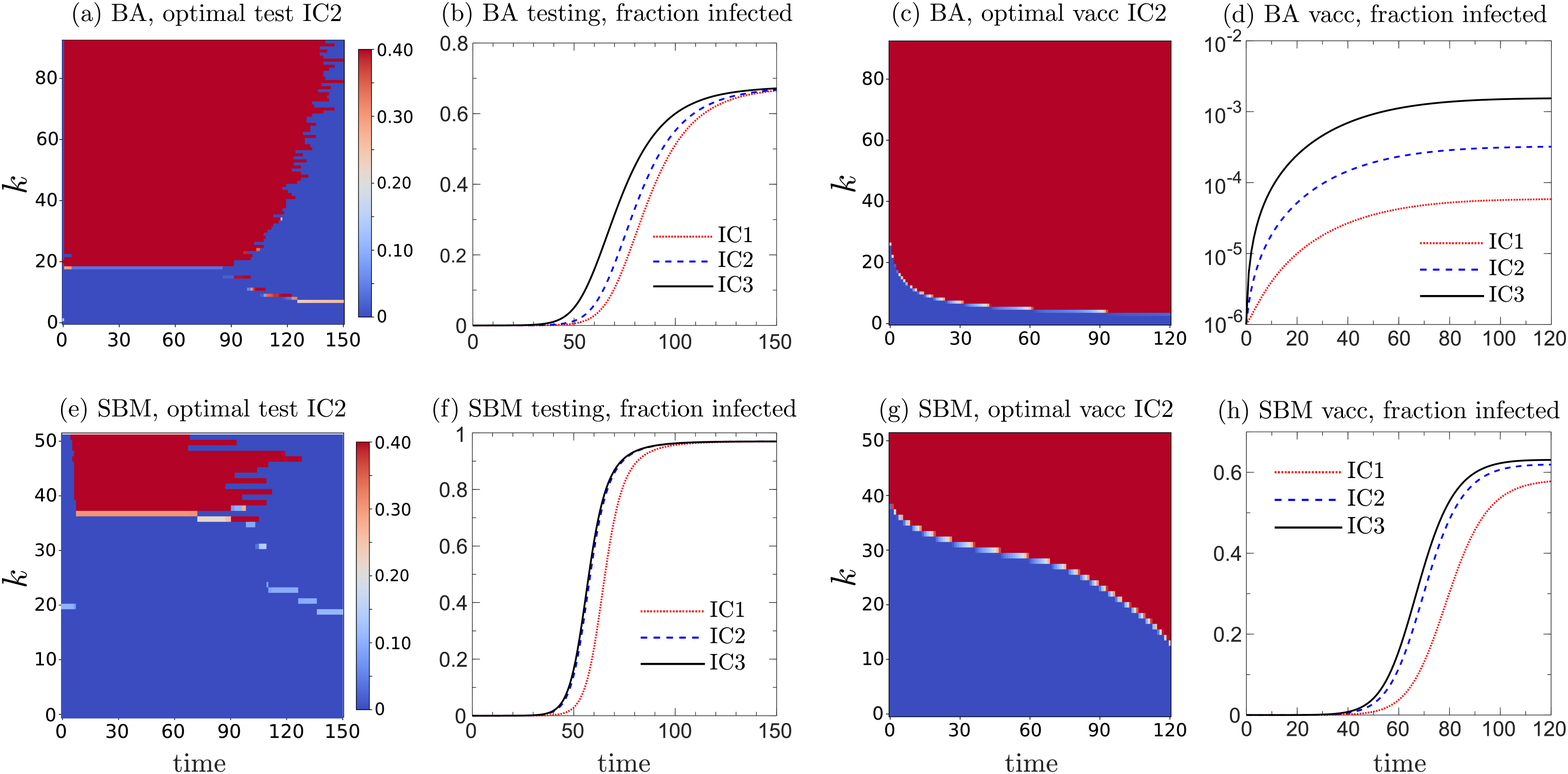}
\end{center}
\vspace{-3mm}
\caption{\small \added{Dependence of intervention effectiveness on the
    degree of the initial infected individual. (a) The PMP-optimal
    testing strategy computed using IC2 ($k_{\rm i} = 20$) on the BA
    network.  Strategies for IC1 ($k_{\rm i}=3$) and IC3 ($k_{\rm
      i}=90$) are qualitatively similar (not shown) with small
    differences at the beginning leading to the different delays in
    the infection dynamics shown in (b). Specifically, for IC1 and IC3
    the initial transient of the optimal testing strategy maximizes
    the testing rate for the subpopulation with the same degree as
    $k_{1}$ and $k_{3}$, respectively, indicating that the optimal
    testing strategy is sensitive to the degree properties initial
    seed infection. Once the disease spreads out, the testing
    strategies ``forget'' the initial condition and converge to each
    other. Despite optimal testing, initial infecteds with higher
    degree, such as IC3, lead to earlier spread of the epidemic.
    Results are found by using a discount factor $\delta = 0.95$, the
    optimal strategy given in Alg.~\ref{algtesting}, and solving
    Eqs.~\eqref{eq:s_k}--\eqref{eq:r_k}.  (c-d) The optimal
    vaccination strategy for IC2 and the associated fraction infected
    for the BA network.  As with testing, the vaccination strategies
    associated with IC1 ($k_{\rm i}=5$) and IC3 ($k_{\rm i}=30$) lead
    to differences in infection magnitudes. However, the optimal
    vaccination strategies are insensitive to different initial
    conditions, even at early times. Since the mechanism of
    vaccination is always to protect high-degree susceptibles, the
    vaccination strategies are not as dependent on the current
    infected population as the testing strategies are.  (e) shows the
    optimal testing strategy for the SBM network, assuming IC2
    ($k_{\rm i}= 20$). (f) The fraction infected exhibits slower
    dynamics for smaller-degree initial conditions. (g) Optimal
    vaccination strategy for IC2 in the SBM network and (h), the
    associated infected fraction showing both delay and amplitude
    changes with changes in the initial condition.}}
\label{fig:ic}
\end{figure*}
\subsection{\added{Dependence on initial conditions}}
\added{Besides the start time of testing or vaccination, initial
  conditions may also affect the optimal strategy. For example, the
  initial propagation of the disease may depend on the degree $k$ of
  the first infected individual \cite{odor2021switchover}.  Instead of
  an initial infectious source who is uniformly distributed across all
  nodes, as described in Eqs.~\eqref{ICs} and \eqref{ICv}, we vary the
  degree of the first infected node and explore how the strategies
  change as a function of concentrated initial condition $i_{k}(0) =
  N_0\delta_{k,k_{\rm i}}/N$. We take $N_0=10^{-6}N$ for both
  networks, $k_{\rm i}=3, 20, 90$ for the BA network, and $k_{\rm i}=
  5, 20, 30$ for the SBM network. These different initial conditions
  are denoted IC1, IC2, and IC3 for each network, respectively.}

\added{Fig.~\ref{fig:ic} shows that under optimal testing and
  vaccination, a smaller degree of the first infected source typically
  leads to a smaller subsequent infected population.  Optimal testing
  strategies are seen to be visibly dependent on the initial
  conditions, \textit{i.e.}, the degree of the initial infected
  patient. This sensitivity arises because testing those who are not
  infected will only waste testing resources. After early stage spread
  of the disease, the testing strategy becomes insensitive initial
  condition because persons with all degrees are infected and those
  with a higher degree tend to be infected sooner.}

 \added{The optimal vaccination strategies obtained through
   Alg.~\ref{algtesting} are also relatively insensitive to initial
   conditions in both networks, particularly at at longer times.
   Although not shown, the optimal strategies associated with
   different ICs are mostly the same times because nodes with larger
   degrees tend to always be vaccinated first to minimize the loss
   function Eq.~\eqref{Lossvaccine}. Susceptibles with higher degrees
   are more vulnerable and should be vaccinated first to mitigate
   subsequent infection events. The strategy differences disappear at
   later times.}

%\subsection{\added{Effects of contact adjusting contact strengths}}
%\added{Finally, we consider transmissibility that depends on the node
%  degree. While network models allow for heterogeneous contacts in a
%  population, an additional important behavioral constraint can
%  incorporated. If human contacts are constrained by time,
%  transportation, infected nodes with very high degree $k$ would not
%  proportionally transmit the disease. In other words, people with a
%  large number of friends would spend proportionally less time with
%  each contact. Thus, the time spent with each friend would scale as
%  $1/k$, and the effective, contact-time-limited transmission rates
%  would take the form $\beta = \beta_{1}/(k\ell)$. Using this form for
%  $\beta$ in \eqref{eq:s_k}, \eqref{eq:i_k}, \eqref{vaccines}, and
%  \eqref{vaccinei}, find the corresponding optimal strategies.  This
%  reduced contact time model suppresses the effective heterogeneity
%  reflected by degree distribution alone. As a consequence, the
%  optimal strategies are closer to those expected by a simple mass
%  action where there is less dependence of $f_{k}$ and $v_{k}$ on $k$.
%  Fig.~\ref{fig:k} shows...}

If more information on contact patterns of individual nodes is
available, it is possible to further refine the proposed policies
using interventions that rely not only on node degrees, but also on
other structural features such as percolation and betweenness
centrality~\cite{newman2018networks,piraveenan2013percolation}.

In addition to the application of control-theoretic methods, we also
utilized reinforcement learning (RL) to identify effective testing and
vaccination strategies. On occasions when the best optimal strategy
can be analytically solved, the controls derived from the Pontryagin's
maximum principle outperform RL-based interventions; yet reinforcement
learning is applicable to epidemic management problems when analytic
solutions are not available. Our results also indicate that
optimal-control theory may be helpful for pre-training and restricting
the space of possible actions, which may lead to more efficient RL
algorithms.

Further generalizations of the derived optimal testing and vaccination
strategies include devising different loss functions other than
Eqs.~\eqref{eq:loss} and \eqref{Lossvaccine} to take into account
factors such as economic effects and prioritizing certain demographics
groups (\eg, individuals with comorbidities). Furthermore, another
possible direction for future research is to devise optimal testing
and vaccination resource allocation strategies under disease-induced
resource constraints~\cite{bottcher2015disease}. Instead of using
RL-based control strategies, it is also a worthwhile direction for
future research to apply neural ODE control
frameworks~\cite{asikis2021neural,bottcher2021implicit} to the studied
resource allocation problems since these control methods showed a
better performance than reinforcement learning and numerical adjoint
system solvers. \added{Finally, in addition to obtaining and even
  combining testing and/or vaccination strategies, our results
  indicate that different network structures (\eg, BA \textit{vs.}
  SBM) have different susceptibilities to optimal strategies. Thus,
  policies such as selective social distances can potentially be used
  to shift network structure towards one that is more sensitive to
  direct testing and vaccination strategies.}

\appendices

\section{Basic reproduction number}
\label{R0}
In this appendix, we analytically derive the basic reproduction number
$\mathcal{R}_0$ for uncorrelated networks and compare the resulting
values with those obtained using Eqs.~\eqref{R0equation} and
\eqref{F_next_gen_matrix}. As a starting point, we note that the
conditional degree distribution $P(\ell|k)$ can be expressed in terms
of a symmetric (for undirected networks) joint degree distribution
$P(\ell,k)$, the probability that a randomly chosen edge connects two
nodes with degrees $\ell$ and $k$.  Marginalizing $P(\ell,k)$ over
$\ell$ yields the distribution over edge
ends~\cite{weber2007generation} $P_{\rm e}(\ell) \equiv \sum_k
P(\ell,k) = \ell P(\ell)/\langle k \rangle$, where $\langle k \rangle
= \sum_k k P(k)$ is the mean degree. The conditional degree
distribution is related to the joint distribution via
\begin{equation}
P(\ell\vert k) = \frac{P(\ell,k)}{P_{\rm e}(k)} = \frac{\langle k\rangle P(\ell, k)}{kP(k)},
\label{eq:correlated}
\end{equation}
which can be further simplified in the uncorrelated network limit 
where $P(\ell,k)\approx P_{\rm e}(k)P_{\rm e}(\ell)$:
\begin{equation}
P(\ell\vert k) \approx \frac{\ell P(\ell)}{\langle k\rangle}.
\label{eq:uncorrelated}
\end{equation}
Eqs. \eqref{eq:correlated} or \eqref{eq:uncorrelated} can be used as a
simpler replacement for $P(\ell|k)$ in Eqs.~\eqref{eq:s_k} and
\eqref{eq:i_k} if $E_{\ell, k}/(kN_{k})$ is not directly accessible.
For example, for an uncorrelated network (\ie, for $P(\ell|k)=\ell
P(\ell)/\langle k\rangle$), we find
\begin{equation}
  \f{\dd i^{\rm u}_k(t)}{\dd t} = \beta^{\rm u} \frac{k s_k(t)}{\langle k\rangle}
  \sum_\ell \ell i^{\rm u}_{\ell}(t)-\gamma^{\rm u} i^{\rm u}_k(t),
\end{equation}
where we have set testing rates $f_k(0)=0$ at the start of the
infection. According to \cite{kiss2006effect}, we define
\begin{equation}
I^{\rm u}(t) \coloneqq \sum_k i_k^{\rm u}(t),\,\,\, J^{\rm u}(t) \coloneqq \sum_k k i_k^{\rm u}(t)
\end{equation}
and obtain
\begin{equation}
\begin{aligned}
\dot{I}^{\rm u}(t) & =\beta^{\rm u} J^{\rm u}(t)-\gamma^{\rm u}I^{\rm u}(t),\\
\dot{J}^{\rm u}(t) & =\beta^{\rm u}
\frac{\langle k^2 \rangle}{\langle k \rangle}J(t)-\gamma^{\rm u}J^{\rm u}(t).
\label{eq:IJ}
\end{aligned}
\end{equation}
We perform a linear stability analysis around the disease-free state
$(I^*,J^*)=(0,0)$ and find the eigenvalues 
to Eqs.~\eqref{eq:IJ}:
\begin{equation}
\lambda_{\pm}=-\gamma^{\rm u}\pm \beta^{\rm u}\frac{\langle k^2 \rangle}{\langle k \rangle}.
\end{equation}

The transition from negative to positive eigenvalues occurs for
$-\gamma^{\rm u}+ \beta^{\rm u} \langle k^2 \rangle/\langle k
\rangle=0$. Hence, the basic reproduction number is
\begin{equation}
\mathcal{R}_0=\frac{\beta^{\rm u}}{\gamma^{\rm u}} 
\frac{\langle k^2 \rangle}{\langle k \rangle}
=\frac{\beta^{\rm u}}{\gamma^{\rm u}}
\left(\langle k\rangle+\frac{{\rm{Var}}[k^2]}{\langle k \rangle}\right).
\label{R0def}
\end{equation}
If we use the conditional degree distribution $P(\ell\vert k)=(\ell-1)
P(\ell)/\langle k\rangle$ proposed by Kiss \textit{et
  al.}~\cite{kiss2006effect} to account for a reduction in neighboring 
susceptible vertices, the corresponding basic reproduction
number is modified to 
\begin{equation}
\mathcal{R}_0^{\rm Kiss}=\frac{\beta^{\rm u}}{\gamma^{\rm u}}
\left(\langle k\rangle-1+\frac{{\rm{Var}}[k^2]}{\langle k \rangle}\right).
\label{R0kiss}
\end{equation}
The mean degrees of the BA and SBM networks are 3.77 and 23.14, and
the variances for the BA and SBM networks are 20.40 and 36.62,
respectively. Using the values $\gamma^{\rm u}=14^{-1}/{\rm day}$
and $\beta^{\rm u}=0.0417/\text{day}$ for the BA network, we find that
the basic reproduction numbers $\mathcal{R}_0=5.361$ and
$\mathcal{R}_0^{\rm Kiss}=4.777$ are larger than 4.5, the value we
used to determine $\beta^{\rm u}$ according to the next-generation
matrix method (Eqs. \eqref{R0equation} and \eqref{F_next_gen_matrix}).
The observed approximation errors in Eqs.~\eqref{R0def} and
\eqref{R0kiss} are a consequence of the assumption that the underlying
network is uncorrelated.  For the SBM network, we find
$\mathcal{R}_0=4.499$ and $\mathcal{R}_0^{\rm Kiss}=4.317$, close to
the 4.5 value used to find $\beta^{\rm u}=0.0130$ using
Eqs.~\eqref{R0equation} and \eqref{F_next_gen_matrix}.

%The value of $\mathcal{R}_0$, calculated
%from Eq.~\eqref{R0def}, is closer to the one obtained with
%Eq.~\eqref{F_next_gen_matrix} than $\mathcal{R}_0^{\text{Kiss}}$

To summarize, our comparison shows that in the SBM model where the
degrees of neighbors are uncorrelated, Eqs.~\eqref{R0def} and
\eqref{R0kiss} give close approximations of the actual reproduction
number calculated from the next-generation matrix method
\eqref{R0equation}. For the BA network, degree correlations make
Eqs.~\eqref{R0def} and \eqref{R0kiss} overestimate the actual
reproduction number. Therefore, we recommend using the next-generation
matrix method to numerically determine the basic reproduction number
unless degree correlations are weak and Eqs.~\eqref{R0def} and
\eqref{R0kiss} can provide accurate estimates of $\mathcal{R}_0$.

\section{Optimal testing and vaccination algorithms}
\label{ALGO}
Below, we explicitly give the pseudo-code for the testing and
quarantine model based on Pontryagin's maximum principle.
\begin{algorithm}
\caption{\small Pseudo-code for determining optimal testing strategies
  based on Pontryagin's maximum principle.}
\begin{algorithmic}[1]
\State Initialize $t=0$, $s_k(0), i^{\text{u}}_k(0), i^*_k(0)$,
$\Delta t$, $T=n\Delta{t}$, $\beta^{\text{u}}, \beta^*$,
$\gamma^{\text{u}}, \gamma^*$, $\delta$, initial strategy
$F(k\Delta{t}), k$, $f_{\max}$, $f_{\min}$, $\epsilon$,
$iter_{\max}$ 
\For{$k=0:n-1$} 
\State Calculate $s_k(t), i_k^*(t),
i_k^{\text{u}}(t)$ under the strategy $F(k\Delta{t})$ from
Eqs.~\eqref{eq:s_k}--\eqref{eq:i_k_ast} 
\EndFor 
\State Set
$\lambda_k^s, \lambda_k^{\text{u}}, \lambda_k^*=0, k=n$ 
\State
Calculate the loss function $L_1$ in Eq.~\eqref{eq:loss}
\For{$k=n-1:0$} 
\State Calculate $\lambda_k^s, \lambda_k^{\text{u}},
\lambda_k^*$ under the strategy $F(k\Delta{t})$ from
Eqs.~\eqref{lambdas}--\eqref{lambdais} 
\EndFor 
\For{$k=0:n-1$} 
\State
First renew the strategy $F(k\Delta{t})$, then calculate $s_k,
i_k^{\text{u}}, i_k^*$ under the strategy $F(k\Delta{t})$ from
Eqs.~\eqref{eq:s_k}--\eqref{eq:i_k_ast} 
\EndFor 
\State Calculate the loss function $L_2$ in Eq.~\eqref{eq:loss}
\State $i\gets1$
\While{$|L_1-L_2|>\epsilon ~\&\&~i< iter_{\max}$} 
\State $i\gets i+1$
\State $L_1\gets L_2$ 
\State Set $k=n, \lambda_k^s, \lambda_k^{\text{u}},\lambda_k^*=0$ 
\For{$k=n-1:0$} 
\State Calculate $\lambda_k^s,
\lambda_k^{\text{u}}, \lambda_k^*$ under the strategy $F(k\Delta{t})$
from Eqs.~\eqref{lambdas}--\eqref{lambdais} 
\EndFor 
\For{$k=0:n-1$}
\State First renew the strategy $F(k\Delta{t})$, then calculate $s_k,
i_k^{\text{u}}, i_k^*$ under the strategy $F(k\Delta{t})$ from
Eqs.~\eqref{eq:s_k}--\eqref{eq:i_k_ast} 
\EndFor 
\State Calculate the Loss function $L_2$ in
Eq.~\eqref{eq:loss} 
\EndWhile
\end{algorithmic}
\label{algtesting}
\end{algorithm}

\newpage
\section{Reinforcement-learning strategy}
\label{sec:rl}
To identify effective testing and vaccination strategies, we also
investigated reinforcement-learning (RL) approaches.  RL explores the
space of all possible actions and directly optimizes the loss
functions for testing and vaccination defined in Eqs.~\eqref{eq:loss}
and \eqref{Lossvaccine}. Here, we use an RL approach with experience
replay to learn both the optimal testing strategy in
Eqs.~\eqref{eq:s_k}--\eqref{eq:r_k} and the optimal vaccination
strategy in Eqs.~\eqref{vaccines}--\eqref{vacciner}. 

Typically, applying a policy-gradient method to a continuous action
space will usually yield poor results due to the inability of such
methods to explore the whole space.  However, using our previous
results based on Pontryagin's maximum principle (PMP), we know that
the optimal strategy is always obtained by maximizing the testing and
vaccination rates for subpopulations presumed to be at a higher risk.

Therefore, we do not need to explore the whole space
%$(\mathbb{R}^+)^{K}$ that contains
of all possible actions. Instead, from Eqs.~\eqref{PMminitest},
\eqref{Hamiltonvaccine}, we can restrict our strategy space to the
extreme points\footnote{Extreme points are points in a set that cannot
  be written as a nontrivial convex linear combination of any other
  points in the same set.} of the set
\begin{equation}
 \{(f_k)|_{k=1}^{K}|\sum_{k=1}^{K}f_k= F(t), f_{\min}\leq \frac{f_k}{N_k}\leq f_{\max}\}
\end{equation}
for determining testing-resource allocation and the extreme points of
the set
\begin{equation} \{(v_k)|_{k=1}^{K}|\sum_{k=1}^{K}v_k
= V(t), v_{\min}\leq \frac{v_k}{Ns_k(t)}\leq v_{\max}\}
\end{equation}
for determining vaccination-resource allocation at each step. The set
of extreme points represents all strategies that maximize the
testing/vaccination rates for some groups and minimize them for other
groups. Such strategies also cannot be written as nontrivial convex
combinations of other strategies. By confining ourselves to extreme
points, the possible action space is reduced to a finite set on which
we perform RL.

Since the curse of dimensionality increases the number of all possible
strategies exponentially with $K$, we further restrict our RL approach
to networks with degree cutoff $K=20$. This additional constraint
allows us to perform RL with a computation time of about \added{30
  days for the testing model on the BA network, 3 days for the testing
  model on the SBM network, 6 hours for the vaccination model on the
  BA network, and 2 hours for the vaccination model on an SBM
  network.}  All computations are performed using Python 3.8.10 on a
laptop with a 4-core Intel(R) Core(TM) i7-8550U CPU @ 1.80 GHz.

%\hl{XX I will double check the running time later and re-run some
%programs. [@Mingtao, can you add some reasonable estimates for
%training times on your machine? Then we should also add your CPU
%type. We can also add my CPU and the training]} days.

\begin{algorithm}
\caption{\small Pseudo-code of Q-Learning in testing resource allocation.}
\begin{algorithmic}[1]
\State Initialize $F , \delta, C$, $i^u_k(0), 
i^*_k( 0), \beta^u, \beta^*, \gamma^u, \gamma^*, M, \epsilon$
\State Initialize replay memory $D$
\State Randomly initialize the hyperparameter $\Theta=\Theta^-$ for
evaluating the action value function $Q^*(\mathcal{S}, \mathcal{A};\Theta)$
\For{\textrm{episode} $\ell=1:M$}
\State Initialize $\mathcal{S}_0$
\For{$t=0:T_{\text{max}}-1$ }
\State With probability $\epsilon$, randomly select an action $a_i$
\State otherwise select $\mathcal{A}_t=\mbox{argmax}_{\mathcal{A}}Q(\mathcal{S}_t, \mathcal{A}; \Theta)$
\State Execute action $\mathcal{A}_t$ and observe reward $R_t$ and state $\mathcal{S}_{t+1}$
\State Store transition $(\mathcal{S}_t, \mathcal{A}_t, R_t, \mathcal{S}_{t+1})$ in $D$
\State Sample random minibatch of transitions $(\mathcal{S}_j, \mathcal{A}_j, R_j, \mathcal{S}_{j+1})$ from $D$ 
\If{$j= T_{\text{max}}-1$ }
\State Set $y_j=R_j$
\Else
\State Set $y_j=R_j + \delta \max_{\mathcal{A}'}\hat{Q}(\mathcal{S}_{j+1}, \mathcal{A}'; \Theta^-)$
\State Perform a gradient descent step on the minibatch $\sum_j
       [y_j-Q(\mathcal{S}_j, \mathcal{A}_j; \Theta)]^2$ with respect to the network
       parameter $\Theta$
\EndIf
\EndFor
\State Every $C$ steps reset $\Theta^-=\Theta$
\EndFor
\end{algorithmic}
\label{algqlearning}
\end{algorithm}
To identify effective testing and vaccination strategies, we use the
reward functions \eqref{eq:loss} and \eqref{Lossvaccine}. We define
the reward at time $t_i=i\Delta{t}$ as
\begin{equation}
R(\mathcal{S}_i, \mathcal{A}_i, i) = \sum_{k=1}^{K}
\left[s_{k}(t_{i+1}) - s_k(t_i)\right],
\end{equation}
the ``negative'' of the number of total infections during the time
period $[t_i, t_{i+1})$. Here, the state $\mathcal{S}_i$ and action
  $\mathcal{A}_i$ are
\begin{equation}
    \begin{aligned}
    \mathcal{S}_i = & (s_1(t_i),\dots,s_{K}(t_i), 
i^{\text{u}}_1(t_i),\dots,i^{\text{u}}_{K}(t_i), \\
\: & \qquad \qquad i^*_1(t_i),\dots, i^*_{K}(t_i))\in (\mathbb{R})^{3K},\\
    \mathcal{A}_i = & (f_1(t_i),\dots,f_{K}(t_i))\in (\mathbb{R})^{K}
    \end{aligned}
\end{equation}
for the testing model Eqs.~\eqref{eq:s_k}--\eqref{eq:r_k} and 
\begin{equation}
    \begin{aligned}
    \mathcal{S}_i & = (s_1(t_i), \dots,s_{K}(t_i), i_1(t_i),
\dots, i_{K}(t_i))\in (\mathbb{R})^{2K},\\
    \mathcal{A}_i & = (v_1(t_i),\dots,v_{K}(t_i))\in (\mathbb{R})^{K}
    \end{aligned}
\end{equation}
for the vaccination model Eqs.~\eqref{vaccines}--\eqref{vacciner}. We
recursively define the state-value function under a certain policy
$\pi$ to be
\begin{equation}
V^{\pi}(\mathcal{S}_i, i)=
\begin{cases}
V^{\pi}(\mathcal{S}_{i+1})\delta + R(\mathcal{S}_i, 
\pi(\mathcal{S}_i)), & t_i < T_{\text{max}},\\
0, & t_i=T_{\text{max}},\\
\end{cases}
\label{vdef}
\end{equation}
where $\pi(\mathcal{S}_i)$ is the action determined under policy $\pi$
given $\mathcal{S}_i$ and $\delta\in(0, 1]$ is a discount factor. We
also define the action-value function to be

\begin{equation}
Q^{\pi}(\mathcal{S}_i, \mathcal{A}_i, i)=
\begin{cases}
V^{\pi}(\mathcal{S}_{i+1})\delta +  R(\mathcal{S}_i, \mathcal{A}_i, i), &t_i<T_{\text{max}}-1,\\
R(\mathcal{S}_i, \mathcal{A}_i, i), &t_i=T_{\text{max}}-1.\\
\end{cases}
\end{equation}
We use $Q^*$ and $V^*$ to denote the action-value and state-value
functions, respectively, under the best policy and apply the deep
Q-learning algorithm, which has been used to find the optimal strategy
\cite{mnih2015human}.

\begin{figure*}[htb!]
        \begin{center}
    \includegraphics[width=6in]{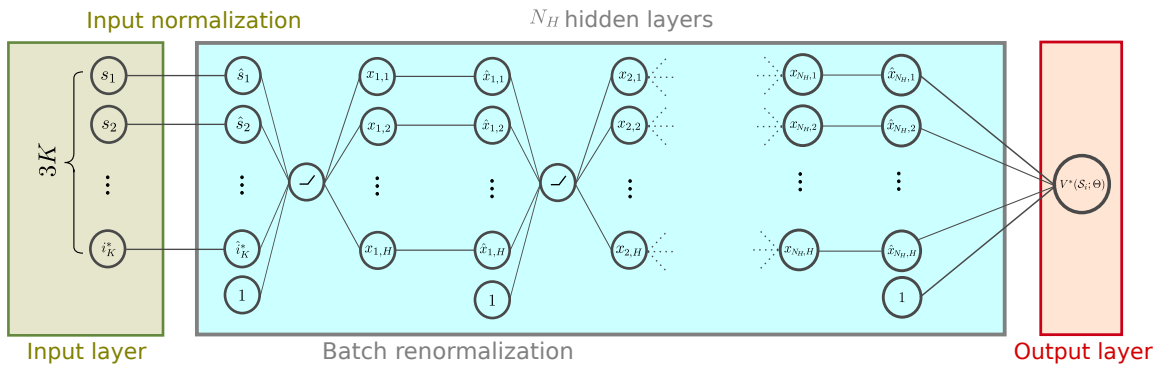}
	\end{center}
\vspace{-3mm}
\caption{\small Illustration of the neural network used to identify
  effective testing and vaccination strategies. The inputs of the
  input layer are $(s_k(t); i^{\text{u}}_k(t);
  i^*_k(t))\in\mathbb{R}^{3K}$. For each hidden layer $i$ ($1\leq i
  \leq N_H$), we normalize the corresponding outputs $x_{i, j}$ for
  all samples in a minibatch such that the resulting values
  $\hat{x}_{i, j}$ have zero mean and unit variance. These values are
  used as inputs to a rectified linear unit (ReLU) activation function
  in the next hidden layer. Neurons labeled 1 are bias terms. The
  output $V^*(S_i; \Theta)$ is an estimate of the state-value function
  under the optimal policy (see Eq.~\eqref{vdef}), where $\Theta$
  denotes the set of hyperparameters.}
%\hl{[LB: I don't think that the hidden layer connections in
      %panel (a) are correct? At least it isn't clear to me how the
      %Batch Renormalization part fits together with the hidden layer
      %illustration/description. Some more comments or changes in the
      %illustration would be necessary. All labels (like input layer,
      %output layer, etc.) should be aligned in the same way (e.g.,
      %centered). And, we can't use $N$ to denote the number of hidden
      %layers because $N$ is the number of nodes. I would suggest using
      %$N_H$.]}}
     \label{neural_network}
\end{figure*}

Here, we use a neural network with a hyperparameter set $\Theta$,
representing neural-network weights to estimate the action-value
function under the best policy $Q^*(\mathcal{S}, \mathcal{A};
\Theta)$, which is improved over epochs by
Alg.~\ref{algqlearning}. \added{We use another neural network with
  hyperparameter set $\Theta^-$ updated every $C=4$ steps to match
  $\Theta$.}
%
%to store the hyperparameter 
%\hl{set $\Theta$ [store  $\Theta$, right?]} for evaluation.} 
%
An illustration of the two neural networks, their layers, and
activation functions is shown in Fig.~\ref{neural_network}.

\added{We use a neural network with $N_H=4$ hidden layers $H=30$
  neurons in each layer. The input data is the state at the
  $i^{\text{th}}$ step $\mathcal{S}_i$, and the output is
  $V^*(\mathcal{S}_i;\Theta)$, the prediction for the optimal
  state-value function generated by the neural network. In each layer,
  the batch normalization technique is used before a rectified linear
  unit (ReLU) function is applied as an activation function. We
  compare the optimal strategies based on the PMP approach from
  Alg.~\ref{algtesting} with Alg.~\ref{algqlearning}. We set $T=100$
  and $\Delta{t}=1$ so that the strategy is updated every day. Here,
  we use $f_{\min} = 0.002/\text{day}, f_{\max} = 0.4/\text{day}$.  We
  used Eq.~\eqref{R0equation} with
  $\gamma^{\text{u}}=(14)^{-1}/\text{day}$ to calculate
  $\beta^{\text{u}}=0.0703/\text{day}$ for the $K=20$ BA network and
  $\beta^{\text{u}}=0.0632/\text{day}$ for the $K=20$ SBM network.}
Both optimal strategies are also compared to the uniform vaccination
strategy \eqref{vaccineunif}. For RL, we train the underlying neural
network for $M=100$ epochs using Alg.~\ref{algqlearning}.
\begin{figure}[htb!]
        \begin{center}
    \includegraphics[width=0.49\textwidth]{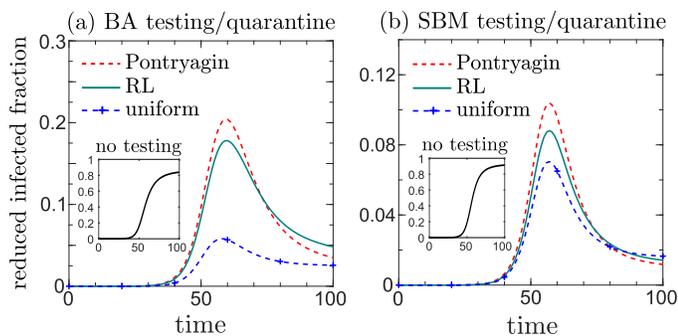}
	\end{center}
\vspace{-3mm}
\caption{\small \added{Reduction in fractions of infected individuals
    calculated as the difference between the fractions infected
    obtained with testing and without testing for the BA network shown
    in (a) and the SBM network shown in (b). The optimal control
    approach based on PMP reduces early infections the most. RL
    outperforms uniform testing in reducing the number of early-stage
    infections. Additionally, the effect of the optimal strategy is
    more striking in the BA network because it has a more
    heterogeneous node degree distribution.}}
     \label{figRLtest}
\end{figure}

Figure \ref{figRLtest} shows the differences between the infected
fractions in simulations with and without testing.  The PMP-based
optimal control reduces early infections the most for both BA and SBM
networks. Early infections contribute more to the loss function
\eqref{eq:loss} since we set the discount factor to $\delta=0.95$. We
also observe that RL-based testing strategies outperform uniform
testing in reducing early-stage infections. Comparing
Fig.~\ref{figRLtest}(a,b), the effect of the optimal vaccination
strategy in the BA network is more pronounced than that in the SBM
network. In the BA network, node degrees are more heterogeneous and
most nodes have small degrees, indicating that epidemic spreading can
be controlled effectively as long as the few high-degree nodes are
monitored and tested. Finally, comparing the result of the
optimal-control approach in Fig.~\ref{figRLtest} with
Fig.~\ref{fig70}, we observe that with a smaller $K$ in the SBM
network, the effect of the optimal vaccination strategy is less
apparent because node degrees are more homogeneous.

Next, we compared the PMP approach with the RL approach for the
optimal vaccination strategy model
Eqs.~\eqref{vaccines}--\eqref{vacciner}.  Here, we set
$v_{\min}=0.0001, v_{\max}=1$.
\begin{figure}
        \begin{center}
    \includegraphics[width=0.49\textwidth]{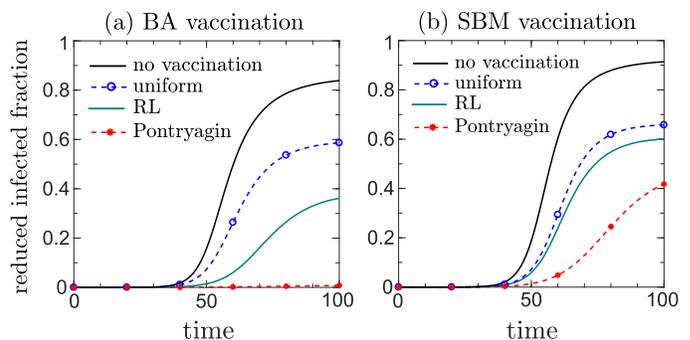}
	\end{center}
\vspace{-3mm}
\caption{\small \added{Reduction in fractions of infected individuals
    calculated as the difference between the fractions infected
    obtained with vaccination and without vaccination for the BA
    network shown in (a) and the SBM network shown in (b).  The
    optimal control approach using PMP can most effectively reduce
    infections for both networks and successfully suppress the
    spreading of the disease in the BA network. On the other hand,
    although not as good as the PMP-optimal strategies, the strategies
    obtained by the RL algorithm Alg.~\ref{algqlearning} can obviously
    reduce infections compared to the uniform vaccination rate
    strategy.  As with testing, we observe that the effect of optimal
    vaccination is more pronounced in the BA network than in the SBM
    network.}}
     \label{fig702}
\end{figure}
For both networks, the optimal vaccination strategy obtained using PMP
can most effectively reduce the initial infections because early
infections have higher weight in the loss function
\eqref{Lossvaccine}. Reinforcement-learning-based vaccination policies
can also reduce initial infections, but the reduction is less than
that of the PMP approach. \added{Comparing Fig.~\ref{fig702}(a,b), we
  again observe that the effect of the optimal vaccination strategy
  for the BA network is more pronounced than that for the SBM network
  because the BA network has a more heterogeneous degree and is
  dominated by small-degree nodes.}

\added{To summarize, the controls derived from PMP are more effective
  than those based on RL. One limitation of RL-based interventions is
  that the possible action space that needs to be explored is usually
  large. However, based on our PMP results, we can constrain the
  action space before the learning process. Such PMP-informed
  constraints allow us to explore just the extreme points of the
  whole action space.  In general, RL could be useful if a procedure
  for computing an explicit solution cannot be formulated.}
%Providing
%  analytical insight into the optimization problem based on optimal
%  control theory could boost training efficiency by ruling out
%  non-optimal actions.}

\newpage

\bibliography{epidemic_control_arXiv.bbl}
\bibliographystyle{IEEEtran}

\begin{IEEEbiography}[{\includegraphics[width=1in,height=1.25in,clip,keepaspectratio]{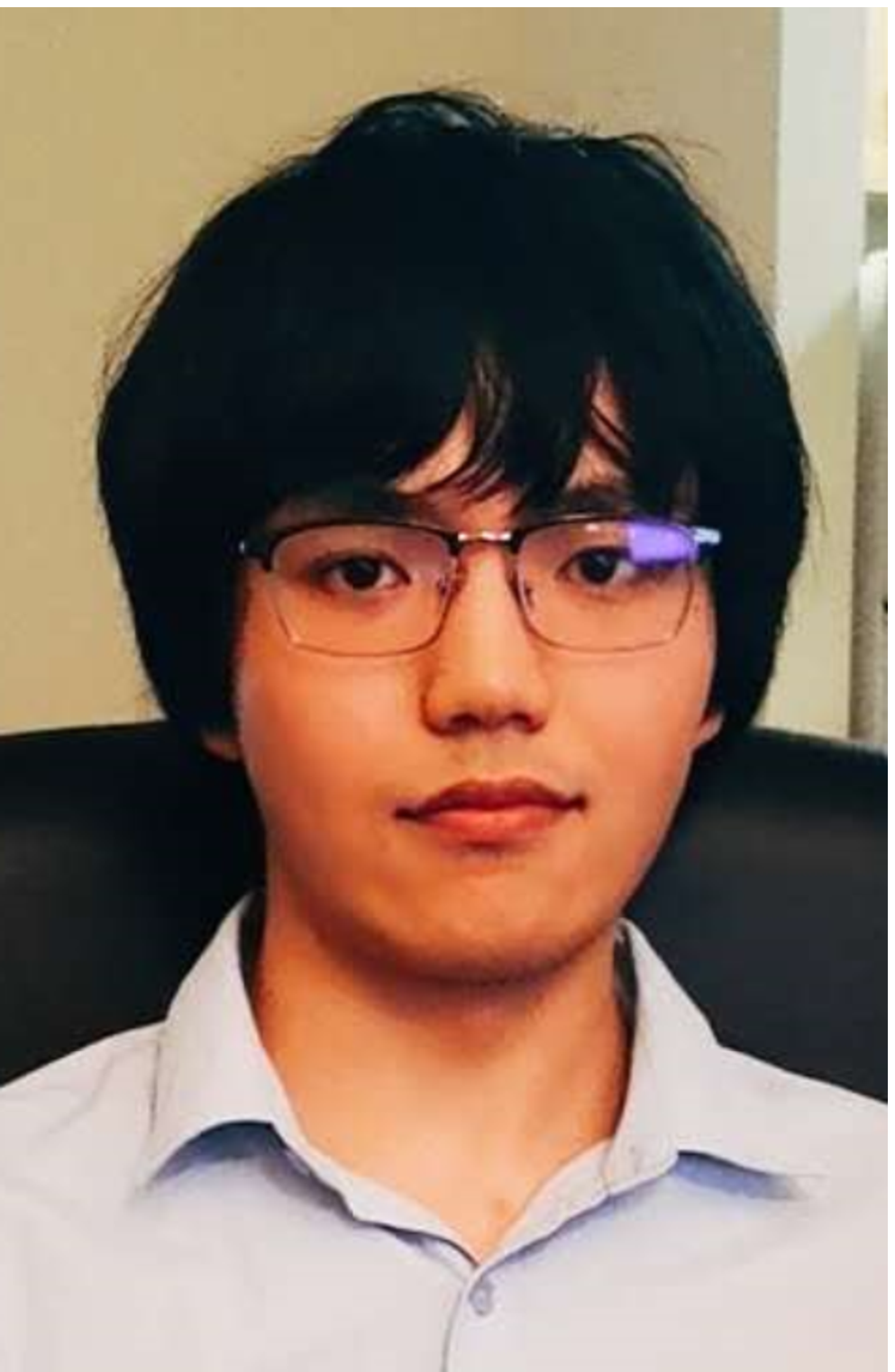}}]{Mingtao Xia}
is a Ph.D. candidate in the Department of Mathematics at the
University of California, Los Angeles. He obtained his bachelors
degree in Information and Computing Science at Peking University in
2019. His research areas include mathematical modeling and
computational methods.
\end{IEEEbiography}
\begin{IEEEbiography}[{\includegraphics[width=1in,height=1.25in,clip,keepaspectratio]{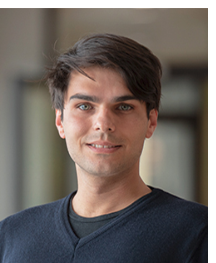}}]{Lucas B\"{o}ttcher} is an assistant professor of Computational Social Science at the
Frankfurt School of Finance \& Management. He completed his doctoral
studies in theoretical physics and applied mathematics at ETH Zurich
in 2018. After working as a lecturer for computational physics at ETH
Zurich, he joined the Dept.~of Computational Medicine at the
University of California, Los Angeles as a fellow of the Swiss
National Fund. His research areas involve applied mathematics,
statistical mechanics, and machine learning.
\end{IEEEbiography}
\begin{IEEEbiography}[{\includegraphics[width=1in,height=1.25in,clip,keepaspectratio]{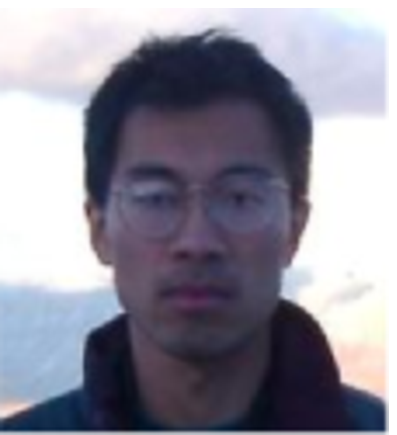}}]{Tom Chou} is a professor in the Departments of Computational Medicine
and Mathematics at the University of California, Los Angeles.  After
obtaining his PhD in physics from Harvard University, he continued
postdoctoral research at Cornell University, the University of
Cambridge, and Stanford University.  His research interests lie in
statistical physics, applied mathematics, and mathematical biology.
\end{IEEEbiography}
\vfill

%\begin{IEEEbiographynophoto}{Lucas B\"{o}ttcher}

%\newpage

%\begin{IEEEbiographynophoto}{Tom Chou}

%%%%%%%%%%%%%%%%%%%%%%%%%%%%%%%%%%%%%%%%%%%%%%%%%%%%%%%%%%%%%%%%%%%%%%%%%%%%%%%%%%%%%%%%%%%%%%
%%%%%%%%%%%%%%%%%%%%%%%%%%%%%%%%%%%%%%%%%%%%%%%%%%%%%%%%%%%%%%%%%%%%%%%%%%%%%%%%%%%%%%%%%%%%%%
\end{document}